\documentstyle[prb,epsfig,aps,multicol,cite]{revtex}
\draft
\author{{R.~Folk$^{(1)}$, Yu.~Holovatch$^{(2,3)}$,
T.~Yavors'kii$^{(3)}$},
\\[1.5ex]
$^{(1)}$Institute for Theoretical Physics,
University of Linz, A-4040 Linz, Austria
\\
$^{(2)}$Institute for Condensed Matter Physics,
Ukrainian Academy  of Sciences, UA-79011 Lviv, Ukraine
\\
$^{(3)}$Ivan Franko National University of Lviv,
UA-79005 Lviv, Ukraine}

\title{
\Large\bf
Effective and asymptotic critical exponents of weakly diluted quenched
Ising model:  3d approach versus $\sqrt{\varepsilon}$--expansion }

\date{October 15, 1999}

\begin{document}

\maketitle
\begin{abstract}
{\small
We present a field-theoretical treatment of the critical behavior of
three-di\-men\-si\-o\-nal weakly diluted quenched Ising model. To
this end we analyse in the replica limit $n \rightarrow 0$ the 5--loop
renormalization group functions of the $\phi^4$--theory with
$O(n)$--symmetric and cubic interactions ({\em H. Kleinert and V.
Schulte-Frohlinde, Phys. Lett. B 342, 284 (1995)}). The minimal
subtraction scheme allows one to develop either the
$\sqrt{\varepsilon}$--expansion series or to proceed within the 3d
approach, performing expansions in terms of renormalized couplings.
Doing so, we compare both perturbation approaches and discuss their
convergence and possible Borel summability.  To study the crossover
effect we calculate the effective critical exponents.
We report resummed numerical values for the
effective and asymptotic critical exponents.  The results obtained within
the 3d approach agree pretty well with recent Monte Carlo
simulations. $\sqrt{\varepsilon}$--expansion does not allow reliable
estimates for $d=3$.
}
\end{abstract}


%{\bf PACS numbers:} 64.60.Ak, 61.43.-j, 11.10.Gh

%%%%%%%%%%%%%%%%%%%%%%%%%%%%%%%%%%%%%%%%%%%%%%%%%%%%%%%%%%%%%%%%%%%%
%                              SECTION I
%%%%%%%%%%%%%%%%%%%%%%%%%%%%%%%%%%%%%%%%%%%%%%%%%%%%%%%%%%%%%%%%%%%%
\section{Introduction \label{I}}

Influence of a weak quenched disorder on magnetic second order phase
transition remains the subject of much interest. One of the central
problems here is the answer to the questions: (i) do the critical
exponents (of a homogeneous magnet) change under dilution (by a
nonmagnetic component)? and, if yes, (ii) how do they change?

Replying the first question it was argued \cite{Harris74} almost
25 years ago that if the heat capacity critical exponent $\alpha$
of the pure (undiluted) system is positive, i.e. the heat
capacity diverges at the critical point, then a quenched
disorder causes changes in the critical exponents. This statement
is known as  Harris criterion. Later, for a large class of
$d$--dimensional disordered systems it was proven that the
correlation length critical exponent $\nu$ must satisfy  the bound
$\nu \geq 2/d$ \cite{Chayes86}. Both statements focus attention
towards studies of the $d=3$ Ising model, where the typical
numerical values of the above exponents, together with the magnetic
susceptibility and the order parameter critical exponents in the pure
case read: $\alpha=0.109 \pm 0.004$, $\nu=0.6304 \pm 0.0013$,
$\gamma=1.2396 \pm 0.0013$, $\beta= 0.3258 \pm 0.0014$ \cite{note1}.

The reply on the second question concerning the numerical values of
the critical exponents of $d=3$ weakly diluted quenched Ising model
(random Ising model - RIM) is more complicated. Below we briefly review
some experimental
\cite{Dunlap81,Birgeneau83,Hastings85,Mitchell86,Barret86,%
Belanger86,Thurston88,Belanger88,Rosov88,Ramos88,Ferreira91,%
Belanger95,Belanger96,Hill97,Slanic98,Slanic98a,Slanic99}
theoretical
\cite{Grinstein76,Khmelnitskii75,Lubensky75,Jayaprakash77,Shalaev77,%
Shalaev97,Folk98a,Folk98,Newman82,Jug83,Holovatch92,Mayer84,%
Holovatch98,Mayer89,Mayer89a,Janssen95}
and numerical
\cite{Landau80,Marro86,Chowdhury86,Braun88,Wang89,Wang90,Holey90,
Heuer90,Heuer93,Prudnikov93,Hennecke93,Wiseman98a,Wiseman98b,Parisi98}
results for the critical exponents of $d=3$ magnets described by the RIM.
Although the statements of Refs. \cite{Harris74,Chayes86} should
hold in principle for  arbitrary weak disorder, the new
critical exponents appear only in a region of temperatures
controlled by the concentration of the nonmagnetic component. Such a
region not always may be reached in practice, where only the effective
exponents are observed \cite{Riedel74}. Theoretical results are due to
the application of the renormalization group approach. These are
numerous for the asymptotic values of critical exponents but quite
seldom for the effective ones \cite{Janssen95}.

This paper has been motivated by recent Monte Carlo calculations of the
asymptotic \cite{Parisi98} and effective
\cite{Heuer93,Wiseman98a,Wiseman98b} critical exponents for the three
dimensional Ising systems with quenched disorder. Here, we deal with a
renormalization group study of both asymptotic and effective critical
behavior. Perturbation theory series, which are known in the 5--loop
approximation for the $O(n)$ model with cubic anisotropy \cite{Kleinert95}
enable us to perform our studies in the replica limit $n \rightarrow 0$.
The minimal subtraction scheme \cite{tHooft72} allows to develop
either the $\sqrt{\varepsilon}$--expansion or to proceed in the 3d
approach, performing expansions in terms of renormalized couplings.
Doing so, we compare both perturbation approaches and discuss their
convergence and possible Borel summability.

The paper is arranged as follows. In  section \ref{II} we give
a brief account of some results on critical properties of systems
of interest, section \ref{III} describes the model and the renormalization
procedure. In section \ref{IV} we analyze the series obtained, perform
their resummation, and give results for the asymptotic values of
critical exponents. Effective critical behavior is discussed in
section \ref{V}. Section \ref{VI} concludes our study. Details of the
resummation procedures exploited in our study are given in the Appendix A.

%%%%%%%%%%%%%%%%%%%%%%%%%%%%%%%%%%%%%%%%%%%%%%%%%%%%%%%%%%%%%%%%%%%%
%                              SECTION II
%%%%%%%%%%%%%%%%%%%%%%%%%%%%%%%%%%%%%%%%%%%%%%%%%%%%%%%%%%%%%%%%%%%%
\section{Review \label{II}}

{\it Experiments.}
Crystalline mixtures of two compounds provide a typical experimental
realizations of a RIM (see Table \ref{table_experiment}).
The first compound is an ``Ising-like'' anisotropic uniaxial
antiferromagnet with dominating short-range interaction (e.g. $FeF_2$,
$MnF_2$), the second one ($ZnF_2$) is nonmagnetic. Mixed crystals
($Fe_xZn_{1-x}F_2$, $Mn_xZn_{1-x}F_2$) can be grown with high
crystalline quality and very small concentration  gradients providing
an excellent realization of random substitutional disorder of magnetic
ions ($Fe^{+2}$, $Mn^{+2}$) by nonmagnetic ones ($Zn^{+2}$). The
experimental evidence of the new critical behavior at a weak
quenched dilution was the nuclear magnetic resonance measurement of the
magnetization in $Mn_{0.864}Zn_{0.136}F_2$ \cite{Dunlap81}. The value
of the magnetisation exponent $\beta$ was found to differ strongly from
that in undiluted sample (see Table \ref{table_experiment} for
details). In a few years this result was corroborated by nuclear
scattering measurements of magnetic susceptibility and correlation
length critical exponents in $Fe_xZn_{1-x}F_2$
\cite{Birgeneau83,Belanger86} and $Mn_xZn_{1-x}F_2$ \cite{Mitchell86}
at different dilutions $1-x$. The linear birefringence measurements
brought about the cusp-like behavior of a specific heat at the
transition point with $\alpha=-0.09 \pm 0.03$ \cite{Birgeneau83}. In
particular this proved that within the error bars the hyperscaling
relation $d\nu + \alpha = 2$ is satisfied.

Crossover phenomena in diluted systems are governed in addition to the
temperature by the concentration of the magnetic component. The
experimentally obtained exponents are often reported to be
effective ones (i.e.  temperature- and dilution-dependent). However,
already in the first experiments on the critical behavior of quenched
diluted Ising-like antiferromagnets it appeared possible to reach the
asymptotic region. Thus,  studying the critical regime in Ref.
\cite{Birgeneau83} the authors found neither a region in reduced temperature
$\tau$ where one finds ``pure'' Ising exponents nor any
evidence of crossover from pure to random exponents. This  was
explained by the crossover  either taking place outside the
critical region or being too slow.  In Ref. \cite{Barret86} the
crossover from pure to diluted critical behavior was studied and a
magnetization exponent $\beta$ which does not change under dilution
for $x \leq 0.05$ was found. The crossover occurs within a very narrow
range of $\tau$  at relatively large values of $\tau$. In Ref.
\cite{Mitchell86} excellent agreement of the measured exponents
$\gamma$ and $\nu$ with the theoretical values of the RIM
was obtained for the temperature range $4\cdot
10^{-4} \leq \tau \leq 10^{-1}$. The value of the critical exponent
$\beta$ was also the subject of a crossover analysis in Refs. \cite{Rosov88,
Thurston88}. Ref. \cite{Thurston88} concludes that the experimental
errors are too large in order to distinguish between the pure Ising model
and the RIM critical behavior. In Ref. \cite{Rosov88} no crossover
was found after the correction-to-scaling has been taken into account,
and the RIM critical behavior was found in the whole temperature range.

It is known that the diluted Ising magnet in a uniform magnetic field
$H$ along the uniaxial direction exhibits static critical behavior of
the random-field  Ising model \cite{Fishman79,Cardy84}. The
random-field Ising model is the subject of intensive recent experimental
studies (see \cite{Belanger91} for a review). Such experiments give
additional information about the critical behavior of the RIM when performed
for $H=0$ \cite{Ramos88,Ferreira91,Belanger95,Belanger96,Hill97,Slanic98,%
Slanic98a,Slanic99}.

{\it Renormalization group results.}
The change of the Ising-type magnets
critical exponents under dilution found its theoretical
confirmation in the
renormalization group (RG) calculations. There, the change of the
universal properties is interpreted as a crossover from the ``pure''
fixed point present in the undiluted model towards the ``random'' one
which characterizes the new critical behavior of the RIM.
The corresponding RG equations are
degenerated on the one loop level, which leads instead of the familiar
expansion in $\varepsilon=4-d$ to the expansion in $\sqrt{\varepsilon}$
\cite{Khmelnitskii75,Lubensky75,Grinstein76}.
This expansion was recently
extended from the 3--loop \cite{Jayaprakash77,Shalaev77} to the 5--loop level
\cite{Shalaev97,Folk98a} for the critical exponents. The
$\sqrt{\varepsilon}$--expansion series for the amplitude ratios
\cite{Newlove83} is available with the 3--loop accuracy as well
\cite{Shpot90}. Although the $\sqrt{\varepsilon}$--expansion  series
are known with pretty high accuracy they seem to be of no use for the
 $d=3$ case \cite{Folk98}.
Alternatively, the scaling field RG approach lead (via the two fixed
points scenario) to the $d=3$ critical exponents values
$\nu=0.70$, $\gamma=1.38$ \cite{Newman82} definitively different from
their ``pure'' counterparts.

RG equations of the massive $d=3$ field theory
\cite{Parisi80} appeared to be the most fruitful tool takling the
problem. The resulting expansions, considered as the asymptotic ones
(see, however, Refs. \cite{Folk98,Bray87,McKane94} and section
\ref{IV} of this paper) were resummed succesively increasing the order
of the perturbation theory series from the 2--loop (2LA)
\cite{Jug83,Holovatch92} through 3--loop (3LA) \cite{Mayer84,Holovatch98}
to the 4--loop (4LA) approximations \cite{Mayer89,Mayer89a}. Depending
on the resummation procedure applied, the results for the
correlation length and the magnetic susceptibility critical exponents
$\{\nu,\gamma\}$ read:  2LA:  \{0.67813, 1.33551\} \cite{Jug83}
\{0.679, 1.337\} \cite{Mayer84}; 3LA: \{0.670, 1.325\} \cite{Mayer84};
\{0.671, 1.328\} \cite{Holovatch98}; 4LA: \{0.6702, 1.3262\}
\cite{Mayer89}; \{0.6680, 1.318\}, \{0.6714, 1.321\} \cite{Mayer89a}.
Critical amplitudes universal ratios at $d=3$  were obtained in 3--loop
\cite{Bervillier92} and 5--loop \cite{Mayer98} approximations.

It is worth to mention here that the renormalization at fixed $d=3$
dimensions does not necessarily mean an implementation of the massive
scheme \cite{Parisi80} in which  most of the theoretical results
\cite{Jug83,Holovatch92,Mayer84,Holovatch98,Mayer89,Mayer89a} were
obtained. Indeed, the minimal subtraction scheme \cite{tHooft72} is also
suited for the $d=3$ renormalization and can be applied without
$\varepsilon$--expansion \cite{Schloms}. Both schemes serve as
complementary tools to obtain information about the critical
behavior of the pure Ising model (or, more generally,
$O(n)$--symmetric $\phi^4$ theory). However in the case of the RIM there
exists only one study relying on the 3d RG approach in minimal
subtraction scheme. In particular the resummed values of the critical
exponents obtained there in a 3--loop approximation read:  $\nu =
0.666$, $\gamma = 1.313$ \cite{Janssen95}.

{\it Griffiths singularities and replica symmetry breaking.}
Although the RG approach does not allow to determine the region of
concentration where the scenario with two fixed points, explained above,
is applicable,
it is generally assumed that it holds at least for weak dilutions.
The idea of RG presumes that one studies the influence of (thermal)
fluctuations around the spatially homogeneous unique ground state. This
holds for the pure system but does not hold for the diluted one. Here, in
a disorder dominated region one finds a macroscopic number of
spatially inhomogeneous ground states, corresponding to the local
minimum solutions of the saddle-point equation for an effective
Lagrangian \cite{Dotsenko95,Dotsenko98}. Physically the inhomogenous
ground state
corresponds to the so-called Griffiths phase \cite{Griffiths69} caused
by the existence of ferromagnetically ordered ``islands'' in the
region of temperatures between the critical temperatures of pure and
diluted systems.  This is described by a replica-symmetry breaking
Lagrangian \cite{Dotsenko95} leading to a  behavior at the
critical point \cite{Dotsenko95a,Wu98} different from the pure one.
The ``traditional'' RG results
\cite{Grinstein76,Khmelnitskii75,Lubensky75,Jayaprakash77,Shalaev77,%
Shpot90,Shalaev97,Folk98a,Folk98,Newman82,Jug83,Holovatch92,Mayer84,%
Holovatch98,Mayer89,Mayer89a,Newlove83,Bervillier92,Mayer98,Janssen95}
are valid for the replica-symmetric Lagrangians \cite{note2}. Since
the usual RG approach is based on an integration over the disorder
at the  begining of the calculation, it cannot give any information
about the region of concentrations where the weak dilution concept holds
\cite{Korzhenevskii98}.

{\it Monte Carlo simulations.}
Monte Carlo (MC) studies of the $d=3$ RIM systems last for almost two
decades \cite{Landau80,Marro86,Chowdhury86,Braun88,Wang89,Wang90,Holey90,
Heuer90,Heuer93,Prudnikov93,Hennecke93,Wiseman98a,Wiseman98b,Parisi98}.
One of the first studies of critical behavior of a RIM
on a simple cubic lattice \cite{Landau80} revealed the
critical exponents in a wide dilution region with no deviations
within the numerical error from the corresponding exponents of the pure
(undiluted) system (see Table \ref{table_MC}).
However these data were objected by MC simulations on larger lattices
\cite{Marro86,Chowdhury86}, where critical exponents
varying continuously with the magnetic sites concentration $x$ were
obtained. Indications of change of the order parameter critical
exponent $\beta$ upon dilution initiated the extension of studies to
determine the other critical exponents and to check the scaling in the
disordered systems. The application of the Swedsen-Wang algorithm to
the $d=3$ RIM \cite{Wang89} resulted in critical exponents for the
susceptibility and correlation length  independent of concentration
over a wide range of dilution.

Due to Refs. \cite{Wang89,Wang90} and especially
\cite{Heuer90,Heuer93} it became clear that the concentration
dependent critical exponents found in MC simulations are  effective
ones, characterizing the approach to the asymptotic region. The effective
exponents $\gamma,\beta$ and $\zeta=1-\beta$ (the last one describes the
divergence of the magnetization-energy correlation function) were
shown \cite{Heuer90} to be concentration dependent in the
concentration region $0.5 \leq x < 1$.  These data were refined
three years later \cite{Heuer93} resulting in more accurate estimates
for the above mentioned exponents and the critical exponent $\nu$ of the
correlation length yielding continuously varying values. The general
conclusion of Refs. \cite{Heuer90,Heuer93} was:  while a simple
crossover between the pure and weakly random fixed point accounts for the
behavior of systems above $x \simeq 0.8$, in more strongly disordered
systems a more refined analysis is needed.  Note that this last statement
is supported by the conjecture of a ''step-like'' universality of the
three dimensional diluted magnets \cite{Prudnikov93}.

The critical behavior of the $d=3$ RIM was reexamined recently
in Refs. \cite{Wiseman98a,Wiseman98b,Parisi98}. In particular,
the simulations \cite{Wiseman98b} revealed that a disorder realized in a
canonical manner (fixing the fraction of magnetic sites) leads to a
different result from those obtained from a disorder realized in a
grand-canonical manner (see Table \ref{table_MC}, where the
values for the second case are denoted by asterisk). The studies of
Ref.  \cite{Parisi98}
were based on the crucial observation that it is important to take into
account the leading correction-to-scaling term in the infinite volume
extrapolation of the MC data.  The simulations confirmed the
universality of the critical exponents of the $d=3$ RIM over a
wide region of concentrations. In particular the value of the
correction-to-scaling exponent $\omega$ was found to be
$\omega=0.37\pm 0.06$ which is almost half as large as the corresponding
value in the pure $d=3$ Ising model $\omega=0.799\pm 0.011$
\cite{note1}. The smallness of $\omega$ in the dilute case explains
its importance for an analysis of the asymptotic critical behavior.

%%%%%%%%%%%%%%%%%%%%%%%%%%%%%%%%%%%%%%%%%%%%%%%%%%%%%%%%%%%%%%%%%%%%
%                              SECTION III
%%%%%%%%%%%%%%%%%%%%%%%%%%%%%%%%%%%%%%%%%%%%%%%%%%%%%%%%%%%%%%%%%%%%
\section{The model and the renormalization group procedure \label{III}}

In this section, we define the RG procedure as well as the main quantities
which we are going to calculate. Making use of the replica method \cite{note2}
and  taking the average over different configurations of quenched disorder
it is possible to show that the RIM critical behavior in the Euclidian
space of $d=4-\varepsilon$ dimensions is governed by an effective Hamiltonian
with two coupling constants \cite{Grinstein76}:
\begin{equation}
\label{1}
{\cal H}(\varphi)=\int {\rm d}^dR \Big\{ {1\over 2} \sum_{\alpha=1}^{n}
\left[|\nabla {\varphi}_\alpha|^2+ m_0^2 {\varphi}_\alpha^2\right]
+
{v_{0}\over
4!} \left(\sum_{\alpha=1}^{n}{\varphi}_\alpha^2 \right)^2 +
 {u_{0}\over 4!} \sum_{\alpha=1}^{n}{\varphi}_\alpha^4 \Big\},
\end{equation}
in the limit $n \rightarrow 0$. Here, $\varphi_\alpha \equiv
\varphi_\alpha (R)$ is the $\alpha$'s replica of a
scalar field; $u_{0}\sim x > 0$, $v_{0} \sim x(x-1)< 0$ is the bare
coupling constant of the fluctuation's  effective interaction
due to the presence of impurities; $m_0$ is a bare mass.

In order to describe the long-distance properties of the model (\ref{1})
in the vicinity of the phase transition point we shall use
the field-theoretical RG approach. The results
for the RG functions
corresponding to (\ref{1}) are obtained on the basis of the dimensional
regularization and the minimal subtraction scheme \cite{tHooft72}.
The renormalized fields, mass and couplings $\phi,m,u,v$
are introduced by:
\begin{eqnarray}
\varphi &=& Z_{\phi}^{1/2} \phi, \nonumber\\
m_0^2 &=& Z_{m^2} m^2 , \nonumber\\
u_0 &=& \mu^{\varepsilon} \frac{Z_{4,u}}{Z_{\phi}^2}u \, , \nonumber\\
v_0 &=& \mu^{\varepsilon} \frac{Z_{4,v}}{Z_{\phi}^2}v \, . \nonumber
\end{eqnarray}
Here, $\mu$ is the external momentum scale
and $Z_{\phi}$, $Z_{m^2}$, $Z_{4,u}$, $Z_{4,v}$ are the
renormalization constants. They are determined by the condition that
all poles at $\varepsilon=0$ are removed from the renormalized vertex
functions.

The RG equations are written bearing in mind that the bare
vertex functions $\Gamma_0^{N}$ are calculated with the help of the
bare Hamiltonian (\ref{1}) as a sum of one-particle irreducible (1PI)
diagrams:
\begin{equation} \label{2}
\Gamma_0^{N} (\{ r \}) =
<\varphi(r_1)\dots \varphi(r_N)>_{\rm 1PI} \, .
\end{equation}
The $\Gamma_0^{N}$ do not depend on the scale $\mu$, and therefore their
derivatives with respect
to $\mu$ at fixed bare parameters are equal to zero. So one gets
\begin{equation} \label{3}
\mu \frac{\partial}{\partial \mu} \Gamma_0^{N}|_0=
\mu \frac{\partial}{\partial \mu} Z_{\phi}^{-N/2}
\Gamma_R^{N}|_0=0 \, ,
\end{equation}
where the index $|_0$ means a differentiation at fixed bare parameters.
Then the RG equation for the renormalized vertex function
$\Gamma_R^{N}$ reads:
\begin{equation} \label{4}
\Big ( \mu \frac{\partial}{\partial \mu} +
\beta_u \frac{\partial}{\partial u} +
\beta_v \frac{\partial}{\partial v} +
\gamma_{m} m  \frac{\partial}{\partial m} -
\frac{N}{2} \gamma_{\phi} \Big )
\Gamma_R^{N}(m,u,v,\mu) =0,
\end{equation}
and the RG functions are given by
\begin{eqnarray}
\beta_u(u,v) &=& \mu \frac{\partial u}{\partial \mu}|_0,
\nonumber \\
\beta_v(u,v) &=& \mu \frac{\partial v}{\partial \mu}|_0,
\nonumber \\
\gamma_{\phi}=2\gamma_2(u,v) &=& \mu \frac{\partial
\ln Z_{\phi}}{\partial \mu}|_0, \nonumber \\
\gamma_m(u,v) &=& \mu \frac{\partial \ln m}{\partial \mu}|_0 =
\frac{1}{2}\mu \frac{\partial \ln Z_{m^2}^{-1}}{\partial \mu}|_0.
\nonumber \end{eqnarray}
Using the method of characteristics the solution of the RG
equation may be written formally as:
\begin{equation}
\Gamma_R^{N}(m,u,v,\mu) = X(\ell)^{N/2}
\Gamma_R^{N}(Y(\ell)m,u(\ell),v(\ell),\mu \ell),
\label{5}
\end{equation}
where the characteristics are the solutions of the ordinary
differential equations (flow equations):
\begin{eqnarray} \nonumber
\ell \frac{d}{d \ell} \ln X(\ell) = \gamma _{\phi} (u(\ell), v(\ell)),
\qquad
\ell \frac{d}{d \ell} \ln Y(\ell) = \gamma _{m} (u(\ell), v(\ell)),
\\
\ell \frac{d}{d \ell} u(\ell) = \beta _{u} (u(\ell), v(\ell)), \qquad
\ell \frac{d}{d \ell} v(\ell) = \beta _{v} (u(\ell), v(\ell)) \qquad
\qquad \label{6}
\end{eqnarray}
with
\begin{equation} \nonumber
X(1)=Y(1)=1, \qquad u(1)=u, \qquad v(1) = v.
\end{equation}
For small values of $\ell$, the equation (\ref{5}) maps the
large length scales (the critical region) to the noncritical point
$\ell=1$.  In this limit the scale-dependent values of the couplings
$u(\ell), \, v(\ell)$ will approach the stable fixed point, provided
such a fixed point exists.

The fixed points $u^*, \, v^*$ of the differential equations
(\ref{6}) are given by the solutions of the system of equations:
\begin{eqnarray} \nonumber
\beta_u(u^*,\, v^*) &=& 0,
\\
\beta_v(u^*, \, v^*) &=&  0.
\label{8}
\end{eqnarray}
The stable fixed point is defined as the fixed point where the
stability matrix
\begin{equation}
B_{ij}= \frac{\partial \beta_{u_i}}{\partial u_j} \, ,
\qquad u_i=\{ u, \, v\}
\label{9}
\end{equation}
possess eigenvalues
$\omega_1, \omega_2$ with positive real
parts.
The stable fixed point, which is reached starting from the initial values
in the limit $\ell\rightarrow 0$, corresponds to the critical
point of the system.
In the limit $\ell \rightarrow 0$ (corresponding to
the limit of an infinite correlation length) the renormalized
couplings reach their fixed point values and
the critical exponents $\eta$ and $\nu$ of the pair correlation
function at $T_c$
and of the correlation length respectively are then given by
\begin{eqnarray}
\eta &=& 2\gamma_2(u^*, v^*),
\nonumber\\
1/\nu &=& 2(1-\gamma_m(u^*, v^*)).
\label{10}
\end{eqnarray}
In the nonasymptotic region deviations from the power laws with
the fixed point values of the critical exponents are governed by
the correction-to-scaling exponent $\omega=min(\omega_1, \omega_2)$ in
accordance with the Wegner expansion \cite{Wegner72}.

The rest of the critical exponents are obtained by familiar scaling
laws. We note, that the expression for correlation length
exponent may be recast in terms of the renormalization constant
$Z_{\phi^2}$ of the two-point vertex function with $\phi^2$ insertion
by a substitution $2\gamma_m=\gamma_{\phi}+\gamma_{\phi^2}$, which
follows from the relations $Z_{m^2}=Z_{\phi^2}Z_{\phi}^{-1}$ and
$\gamma_{\phi^2}=\mu {\partial \ln Z_{\phi^2}^{-1}}/{\partial
\mu}|_0$.

%%%%%%%%%%%%%%%%%%%%%%%%%%%%%%%%%%%%%%%%%%%%%%%%%%%%%%%%%%%%%%%%%%%%
%                              SECTION IV
%%%%%%%%%%%%%%%%%%%%%%%%%%%%%%%%%%%%%%%%%%%%%%%%%%%%%%%%%%%%%%%%%%%%
\section{The resummation and the results \label{IV}}

In this chapter we analyse series for the RIM
$\beta$-functions and critical exponents.
The RG functions of the corresponding effective Hamiltonian (\ref{1})
in the replica limit $n=0$ have been obtained
up to 5--loop order from the appropriate expressions
for RG functions of the $\phi^4$--theory with $O(n)$--symmetric and
cubic interactions \cite{Kleinert95} and read:
\begin{eqnarray}
\beta_u/u &=& - \varepsilon  + 3u + 4v - 17/3u^2 - 46/3vu - 82/9v^2
\nonumber\\
&+& 32.54968284u^3 + 123.1987313vu^2 + 158.1816418v^2u +
60.32526811v^3 \nonumber\\
&-& 271.6057842u^4 - 1318.116311vu^3 - 2452.429994v^2u^2 -
2003.560971v^3u \nonumber\\
&-& 559.7143854v^4 + 2848.568254u^5 + 16789.89843vu^4 +
40367.08593v^2u^3 \nonumber\\
&+& 48971.12730v^3u^2 + 29091.77179v^4u + 6377.751189v^5,
\label{11}\\
\beta_v/v &=& - \varepsilon  + 8/3v + 2u - 14/3v^2 - 22/3vu - 5/3u^2
\nonumber\\
&+& 25.45714897v^3 + 62.25499170v^2u + 36.36645522vu^2 + 7u^3
\nonumber\\
&-& 200.9263690v^4 - 667.3761895v^3u - 650.5641816v^2u^2 -
259.2586891vu^3 \nonumber\\
&-& 39.91261012u^4 + 2003.976188v^5 + 8469.158907v^4u +
11721.60876v^3u^{2} \nonumber\\
&+& 7434.635066v^2u^3 + 2344.277996vu^4 + 301.5110976u^5,
\label{12}\\
\gamma_2 &=&
1/18v^{2} + 1/6vu + 1/12u^{2} - 1/27v^{3} - 1/6v^{2}u - 3/16vu^{2} -
1/16u^{3}
\nonumber\\
&+& 125/648v^{4} + 125/108v^{3}u + 145/72v^{2}u^{2} + 65/48vu^{3} +
65/192u^{4}
\nonumber\\
&-& 1.005978154v^{5} - 7.544836154v^{4}u - 18.04854621v^{3}u^{2}
\nonumber\\
&-& 19.07838990v^{2}u^{3} - 9.627924878vu^{4} - 1.925584976u^{5},
\label{13}
\\
\gamma_m &=&
 1/3v + 1/2u - 5/18v^{2} - 5/6vu - 5/12u^{2} + 37/36v^{3} + 37/8v^{2}u
\nonumber\\
&+& 251/48vu^{2} + 7/4u^{3} - 5.3808017v^{4} - 32.28481v^{3}u -
57.177011v^{2}u^{2}
\nonumber\\
&-& 39.765731vu^{3} - 9.97815253u^{4} + 37.850485v^{5}
\nonumber\\
&+& 283.878638v^{4}u + 686.375317v^{3}u^{2} + 737.493196v^{2}u^{3}
\nonumber\\
&+& 376.1776339vu^{4} + 75.37777445u^{5}. \label{14}
\end{eqnarray}

In order to obtain the qualitative characteristics of the RIM
critical behavior one
can proceed in two ways. The historically first scheme is known as
$\varepsilon$--expansion and consists (i) in expanding the values of
the couplings at the stable fixed point in $\varepsilon$, (ii)
inserting these expansions into the field theoretic functions for the exponents
(\ref{13}), (\ref{14}) and (iii) expanding again in $\varepsilon$.
Due to the degeneracy of the $\beta$--functions (\ref{11}), (\ref{12})
at 1--loop level, an expansion in $\sqrt{\varepsilon}$ has to be performed
\cite{Khmelnitskii75,Lubensky75,Grinstein76}. Then in 5--loop order
on the basis of Eqs. (\ref{11})--(\ref{14}) one obtains \cite{Shalaev97,Folk98a}:
\begin{eqnarray}
\nu  &=& 1/2 + 0.08411582\varepsilon^{1/2}  - 0.01663203\varepsilon  +
0.04775351\varepsilon ^{3/2} + 0.27258431\varepsilon^2,
\nonumber\\
\eta  &=&  - 0.00943396\varepsilon  + 0.03494350\varepsilon ^{3/2} -
0.04486498\varepsilon ^{2} + 0.02157321\varepsilon ^{5/2},
\label{15} \\
\gamma  &=& 1 + 0.16823164\varepsilon^{1/2} - 0.02854708\varepsilon  +
0.07882881\varepsilon ^{3/2} + 0.56450490\varepsilon ^{2},
\nonumber \\
\omega_1&=&2\, \varepsilon+3.704011194 \,
\varepsilon^{3/2}+11.30873837 \,\varepsilon^2 \, ,
\label{16} \\
\omega_2&=&0.6729265850\, \varepsilon^{1/2}-1.925509085 \,
\varepsilon
-0.5725251806 \, \varepsilon^{3/2} -13.93125952 \, \varepsilon^2 \, .
\nonumber
\end{eqnarray}
Another method consists in (i) fixing the
value of $\varepsilon$ i. e. the lattice dimensionality, (ii) solving the
system of equation for the fixed point and (iii) substituting the fixed
point values of the couplings into  the series for the critical exponents
\cite{Schloms} (the so-called 3d approach).

One should note here that often, for the sake of convenience, within the
3d as well as the $\varepsilon$ expansion one deals with the
expansions of some combinations of the critical exponents instead of
working with them directly on the basis of expressions (\ref{10}).  In
the present paper the values of critical exponents are calculated from
the expansion for $1/\nu-1=1-2\gamma_m$ and the inverse exponent for
magnetic susceptibility $\gamma^{-1}=(1-\gamma_m)/(1-\gamma_2)$.  The
numerical values of the other exponents are obtained by the familiar
scaling laws.

It is well known that the perturbation theory series  for the
RG functions in the weak coupling limit as well as  in
$\varepsilon$--expansion are asymptotic at best. In order to compare
the results obtained on the basis of the $\sqrt\varepsilon$ expansion
and of the 3d approach we have to refer to resummation procedures  in the
calculation of critical exponents. Adjusting the resummation procedure we
discuss first the one--variable case in  both schemes. We start
from the $\varepsilon$--expansion of the pure Ising model critical
exponents which in the 5--loop approximation reads \cite{Kleinert91}:
\begin{eqnarray}
\nu &=& 1/2 + 1/12\varepsilon +
0.04320988\varepsilon^{2} - 0.01904337\varepsilon^{3}
+ 0.07088376\varepsilon^{4} - 0.21701787\varepsilon^{5},
\nonumber\\
\gamma &=& 1 + 1/6\varepsilon  +
0.07716049\varepsilon^{2} - 0.04897495\varepsilon^{3}
+ 0.14357422\varepsilon ^{4} - 0.44662483\varepsilon^{5}.
\label{17}
\end{eqnarray}

An analysis of the $\varepsilon$--expansion case
starts by representing the expressions for the critical
exponents $\nu$ and $\gamma$ of the pure Ising model (\ref{17})
in the form of the Pad\'e approximant:
$\left[ M/N \right](x)=\sum_{i=0}^M a_i x^i/\sum_{j=0}^N b_j x^j$ in
the variable $x=\varepsilon$. The results are shown in the form of a
Pad\'e table (Table \ref{tab3}). The number of the row, $N$, and of
the column, $M$, corresponds to the order of the numerator and the
denominator of the
Pad\'e approximant $\left[ M/N\right]$ respectively. One can see from
this table the expected convergence of the values in the diagonal and
first off--diagonal. Therefrom we estimate the values of the critical
exponents to be $\nu=0.628, \gamma=1.236$. These values can be
compared with the most accurate values $\nu=0.628 \pm 0.001,
\gamma=1.234 \pm 0.002$ \cite{Gorishny84}, obtained by means of more
sophisticated resummation procedures  \cite{Vladimirov79} on the basis
of the 5--loop $\varepsilon$--expansion. We conclude that this good
agreement justifies the application of the Pad\'e analysis for the
$\varepsilon$--expansion series (\ref{17}) \cite{note}.

Possessing the information on the asymptotic divergence of the
$\varepsilon$--expansion we apply a more complicated Pad\'e--Borel
resummation technique which takes into account the factorial
divergence of the series terms. The resummation procedure consists in
several steps:
\begin{itemize}
\item
starting from the initial sum $S$ of $L$ terms one
constructs its Borel--Leroy image:
\begin{equation} \label{18}
S=\sum_{i=0}^La_ix^i\Rightarrow \sum_{i=0}^L\frac{a_i(xt)^i}
{\Gamma(i+p+1)}, \end{equation}
where $\Gamma(x)$ is the Euler's gamma function and $p$ is an
arbitrary nonnegative number;
\item the Borel--Leroy image (\ref{18}) is
extrapolated by a rational Pad\'e approximant
$$\left[ M/N \right](xt);$$
\item
the resummed function $S^{res}$ is obtained in the form:
\begin{equation} \label{Pade-Borel}
S^{res}=\int_0^\infty dt \exp (-t)t^p\left[ M/N \right]
(xt).
\end{equation}
\end{itemize}

The values of the critical exponents $\nu$  and $\gamma$ obtained with
the Pad\'e--Borel resummation for different $M$ and $N$ are
given in  Table \ref{tab4}.  As far as the Pad\'e approximant
enters  the integral, it might happen that the
integrand  contains poles. If this is the case, the corresponding
number in Table \ref{tab4} is calculated  by analytic continuation
taking the principal values of the integral. For the sake of
completeness we include such numbers in Table \ref{tab4}
as well as in the forthcoming Table \ref{tab7}. However, the final
results will be displayed on the basis of data which did not require
such analytic continuation. Except from the $\left[ 1/1 \right]$ case
all our final results
 are obtained from  approximants with a linear denominator
(second columns), which reconstitutes the sign--alternating behavior
of the initial series (\ref{17}).  For the
Borel resummed Pad\'e approximant $\left[ 4/1 \right]$ the estimates
for the exponents $\nu=0.629, \gamma=1.236$ are in a good agreement
both with the above Pad\'e analysis (Table \ref{tab3}) as
well as with the  data of Ref. \cite{Gorishny84} given above.

In order to complete the study of the pure Ising model we perform an
analysis based on the 3d approach. We resum the corresponding RG
functions of the pure model (they can be obtained by putting $v=0$ in
the diluted model RG functions (\ref{11})--(\ref{14})) by means of the
Pad\'e--Borel resummation technique with a linear denominator
approximant (see Appendix A).  The results obtained with this method
are shown in Table \ref{tab5}. One should compare them with the
results obtained recently from the RG--functions in the
3d massive field--theoretical approach, $\nu=0.6304 \pm 0.0013$,
$\gamma=1.2396 \pm 0.0013$ \cite{note1}, and the results of 3d minimal
subtraction scheme in the 5--loop approximation $\nu=0.629 \pm 0.005,
\gamma=1.235 \pm 0.005$ \cite{Schloms}. Comparing all the results
for the pure Ising model we conclude, (i) that the
application of Pad\'e--Borel resummation technique yields accurate
results and, (ii) both the the $\varepsilon$--expansion and 3d
approach lead to reliable values for the exponents.

We now turn to the results of the Pad\'e and the Pad\'e--Borel
resummation techniques applied to the
$\sqrt{\varepsilon}$--expansions (\ref{15})--(\ref{16}) and
the 3d approach RG functions (\ref{11})--(\ref{14})
of the weakly diluted Ising model. We construct Pad\'e approximants
and perform the Pad\'e--Borel resummation introducing an auxiliary variable
$t$ in the expressions (\ref{15})--(\ref{16})  by the substitution:
$\varepsilon \rightarrow \varepsilon t^2$ and putting $t=1$ in
the final results.

The appropriate values for the $\sqrt{\varepsilon}$--expansions are
listed in Tables \ref{tab6} and \ref{tab7} in the same notations as
in the Tables \ref{tab3} and \ref{tab4}.  One can see from the tables
that neither the experimental and nor the reliable theoretical values
listed in Section \ref{II} are obtained.  Moreover, considering
the expansions for the stability matrix eigenvalues (\ref{16})
it turns out that no stable fixed point  exists  in a strict
$\sqrt\varepsilon$--expansion,  even with resummation.

Comparing with the corresponding data for the pure Ising model
we conclude that the nature of the $\sqrt{\varepsilon}$--expansion
series does not allow to obtain reliable information at $d=3$ by means
of the methods mentioned for the case of pure Ising model.  This can
be considered as an indirect evidence of the nonasymptotic nature
of the $\sqrt{\varepsilon}$--expansion.  Thus a different kind of
analysis for the $\sqrt{\varepsilon}$--expansion has to be developed.
The fact that $\varepsilon$-expansion will not be able to give
information on critical exponents in system with quenched disorder was
predicted already in Refs. \cite{Bray87,McKane94}. There, studying the
randomly diluted model in zero dimensions, it was shown that the
non-Borel summable properties of the perturbation theory  series are
the direct consequence of the existence of Griffiths-like singularities
\cite{Griffiths69} caused by the zeroes of the partition function of the
pure system.

In order to treat the theory directly at $d=3$ we need a
generalisation of the Pad\'e--Borel resummation technique to the case of
two variables since the RG functions of the weakly diluted Ising model
depend on two couplings. The corresponding Chisholm--Borel resummation
technique can be defined as follows \cite{Jug83}:
\begin{itemize}
\item
constructing the Borel--Leroy image of the initial $L$th order polynomial
$S$ in the variables $u$ and $v$:
\begin{equation} \label{20}
S=\sum_{0\leq i+j \leq L} a_{i,j}u^iv^j\Rightarrow
\sum_{0\leq i+j \leq L}\frac{a_{i,j}(ut)^i(vt)^j}{\Gamma(i+j+p+1)},
\end{equation}
where $\Gamma(x)$ is the Euler's gamma function and $p$ is an
arbitrary nonnegative number;
\item
extrapolating the Borel--Leroy image (\ref{20}) by a rational
Chisholm approximant \cite{Chisholm73}
$\left[ M/N \right](ut,vt)$ which can be defined as a
ratio of two polynomials both in variables $u$ and $v$,
of degree $M$ and $N$ so that the first terms of its expansion are
equal to those of the function which is approximated;
\item
the resummed function $S^{res}$ then reads:
\begin{equation} \label{21}
S^{res}=\int_0^\infty dt
\exp (-t)t^p\left[ M/N \right] (ut,vt).
\end{equation}
\end{itemize}
Here, similarly to the pure case we restrict the approximants to
linear denominators and choose the value of the fitting parameter $p=0$.
The motivation of such a choice is discussed in detail in Appendix
A.

\begin{figure}[htbp]
\begin{centering}
\setlength{\unitlength}{1mm}
\begin{picture}(128,130)
\epsfxsize=140mm
\epsfysize=100mm
%\put(0,0)
{\epsffile[7 8 832 584]{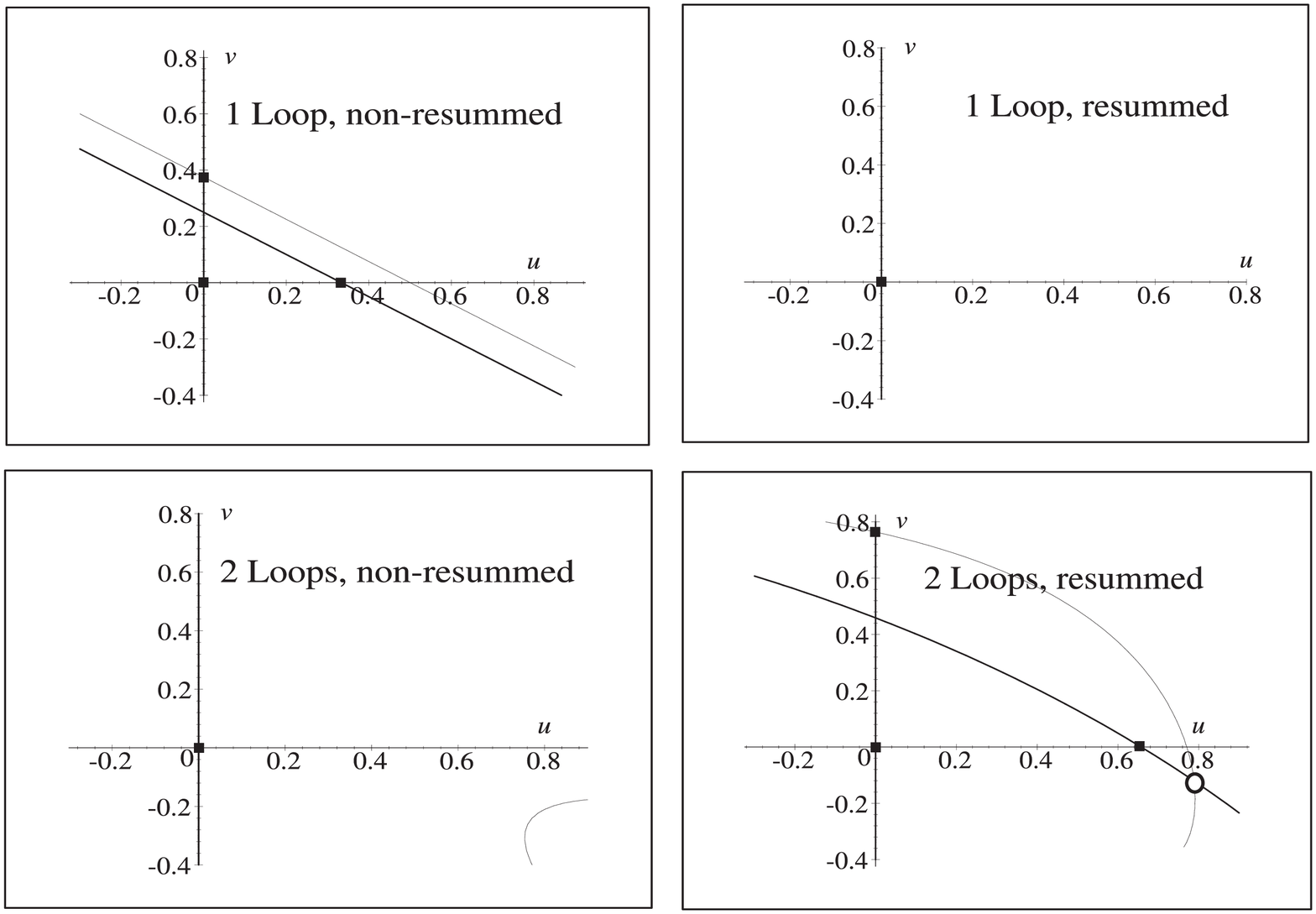}}
\end{picture}
\end{centering}
\caption{\label{comp1}
The lines of zeros of nonresummed (left-hand column) and resummed
by the Chi\-sholm-Borel method (right-hand column) $\beta$--functions
in different orders of the perturbation theory:  1-- and 2--loop
approximations. Thick line corresponds to $\beta_u=0$, thin line
depicts $\beta_v=0$. One can see the appearance of the random fixed
point $u^*>0, v^*<0$ in the 2--loop approximation for the resummed
$\beta$--functions. Stable fixed point is shown by an open circle,
unstable ones are shown by filled boxes.} \end{figure}

\begin{figure}[htbp]
\begin{centering}
\setlength{\unitlength}{1mm}
\begin{picture}(128,130)
\epsfxsize=140mm
\epsfysize=100mm
%\put(-10,0)
{\epsffile[7 8 832 584]{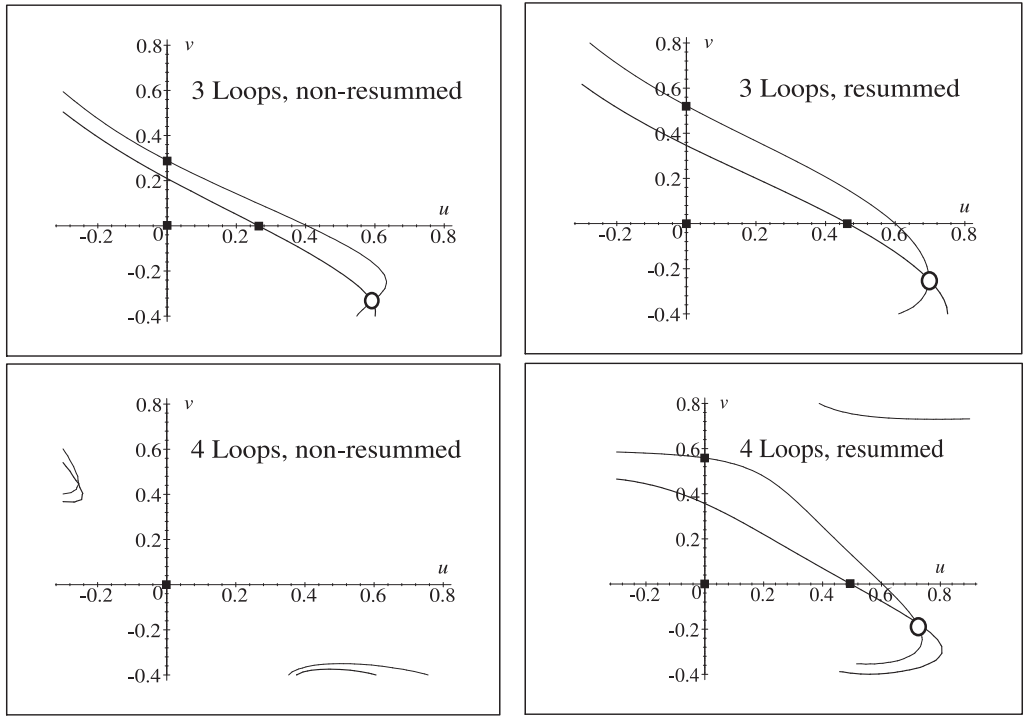}}
\end{picture}
\end{centering}
\caption{\protect\label{comp2}
The lines of zeros of nonresummed (left-hand column) and resummed by
the Chi\-sholm-Borel method (right-hand column) $\beta$--functions in
3-- and 4--loop approximations. The notations are the same as in Fig.
\protect\ref{comp1}. Close to the random fixed point the behavior of
the resummed functions remains alike with the increase of the order of
approximation. This is not the case for nonresummed functions.  }
\end{figure}

Treating the $\beta$-functions by means of this resummation technique
leads to a random fixed point, $u^*>0, v^*<0$, of the model already
in the 2--loop approximation.  The stability analysis shows that this
fixed point is stable proving the crossover to a new critical
regime under dilution.  In Figs. \ref{comp1}, \ref{comp2} we show
the curves $\beta_u(u,v)=0$, $\beta_v(u,v)=0$ in
the $u-v$ plane. The intersections of these curves (i.e. simultaneous
zeros of both $\beta$--functions) correspond to the fixed points, the
stable and unstable points are marked by open circles and filled
boxes respectively. The ``naive'' analysis of the $\beta$--functions,
without applying any resummation procedure leads to the curves,
which are shown on the left-hand side of the Figs. \ref{comp1},
\ref{comp2}. Without resummation only in the 3--loop approximation
one gets stable a random fixed point $u^*\neq0, v^*\neq0$. However the
fixed point disappears in the 4--loop approximation. A completely
different picture is observed when
the resummation procedure is applied (right hand columns of the
figures). In the region of interest for the values of the couplings
the topology of the
fixed point picture remains stable increasing the order of
approximation from the 2--loop to the 4--loop level; the same behavior
has been observed \cite{Holovatch98} for the $\beta$--functions
obtained in the massive 3d scheme \cite{Mayer89}.

The corresponding values of the random fixed point coordinates,
critical exponents and eigenvalues of the stability matrix are listed
in the Table \ref{tab8}. One can see that increasing the order of
approximation one reaches convergent results, compatible with
experimental and theoretical data (see sec. \ref{II}).

Considering the estimation of the accuracy of the results we note
that setting $v=0$ in the RG functions (\ref{11})--(\ref{14}) they are
transformed into the appropriate functions of the pure Ising model. In
this case the deviation of our 4--loop result (see Table \ref{tab5})
from the most accurate one obtained within massive field RG scheme on
the basis of 6--loop expansion for $\beta$--functions and 7--loop
expansion for $\gamma$--functions \cite{Guida98} is within parts of a
percent. This can be considered as a upper bound for the accuracy of RIM
results. Here the comparison of our data with the results of 4--loop
massive scheme results \cite{Mayer89,Mayer89a} yields an
accuracy of several percents. Since the series for $\beta$-- and
$\gamma$--function (\ref{11})--(\ref{14}) are sign--alternating, the
unknown exact stable point coordinates and critical exponents must lie
between the 3-- and 4--loop values giving the same estimate (see
Appendix A about adjusting a free parameter for the fastest
convergence of results). Thus, we conclude the accuracy of the RIM critical
exponents obtained in our study to be of order of several percents.

A peculiarity of the Table \ref{tab8} is that in 5--loop order
the applied resummation technique does not lead to a real root for the
random fixed point. It is expected that the applications of more
sophisticated resummation procedures incorporating the higher order
behavior, still unavailable, will permit an improvement of the estimates
for the critical exponents in the 2--, 3-- and 4--loop level as well as to
obtain them in the 5--loop level.  However, it is not excluded that the
absence of a fixed point solution on the 5--loop level might be connected with
a (possible) Borel-nonsummability of the series under consideration.
In this case the 4--loop approximation will be an ``optimal
truncation'' for the resummed perturbation theory series, similar to
the nonresummed asymptotic series, e.g. in the
$\varepsilon$--expansion of $O(n)$--symmetric $\phi^4$ model, where
``naive'' interpretation of the series truncated by $\varepsilon^2$
term leads to the best (optimal) result.  On the other hand, numerical
and analytic studies of a toy model of a disordered system
\cite{Bray87,McKane94} revealed two possible regimes of the high-order
behavior: the first one corresponds to a Borel-summable series whereas
the second one does not correspond to a Borel-summable series. Numerical
studies of up to 200 terms of the expansion \cite{Bray87} resulted in
the conclusion, that the convergence of the Borel-resummed results
depends on the strength of disorder (relation of the couplings $u/v$).
The convergence of the numerical data of  Table \ref{tab8} is
evidence of the fact that the fixed point values $u^*,v^*$ in $d=3$
lie in a region, where the resummed series gives reliable information.
In any case,  on the basis of the above analysis one can definitely say:
while for the pure Ising model the $\varepsilon$--expansion and
the 3d approach analysis of the RG functions are of equal usefulness,
an interpretation of these functions in the diluted model can be done
only within the framework of the 3d approach.
The application of the $\sqrt{\varepsilon}$--expansion remains to be
valid only near the upper critical dimension $d=4$. This conclusion
holds at least within the discussed resummation schemes.

%%%%%%%%%%%%%%%%%%%%%%%%%%%%%%%%%%%%%%%%%%%%%%%%%%%%%%%%%%%%%%%%%%%%
%                              SECTION V
%%%%%%%%%%%%%%%%%%%%%%%%%%%%%%%%%%%%%%%%%%%%%%%%%%%%%%%%%%%%%%%%%%%%
\section{Effective critical behavior \label{V}}

If one is not within the asymptotic region of the stable fixed point
power laws for the physical quantities may only apply with
effective exponents. The critical behavior is then to be understood as
a crossover behavior between the uncritical background
behavior and the true asymptotic behavior. As it has been noted in
Ref. \cite{Janssen95} this has in principle nothing to do with
crossovers between special fixed points (e.g. the pure one and the
random one).  However, depending on the nonuniversal parameters
entering the nonasymptotic behavior, the crossover may be more or
less influenced by the unstable pure fixed point.

The effective exponents are defined by logarithmic derivatives of
corresponding thermodynamical quantities with respect to reduced
temperature $\tau$ \cite{Riedel74}. In the RG scheme they are
calculated in the region where couplings have not yet reached their
asymptotic (fixed point) values and depend on the flow parameter
$\ell$. For instance the magnetic susceptibility effective exponent
$\gamma_{eff}$ is defined by:  \begin{equation} \label{expeff}
\gamma_{eff}(\tau) = \frac{{\rm d}\ln \chi(\tau)}{{\rm d}\ln t} =
\gamma(u(\ell(\tau)), v(\ell(\tau)))+ \dots, \end{equation} where the
second part is proportional to the $\beta$--functions and comes from
the change of the amplitude part of the susceptibility. In order to
proceed we have to neglect this part since we do not know the
amplitude function in the same orders as the field-theoretical
functions for the exponents. Moreover, the contribution of the
amplitude function to the crossover seems to be small, at least in
other cases \cite{Nasser95,Frey90}. This approximation has also been
used in the earlier work on diluted models  \cite{Janssen95}.

The flow parameter $\ell(\tau)$ may be related to the temperature  via
the matching condition $m(\ell)=(\xi_0^{-1}\ell)^2$ and $m_0^2\sim
\tau$.  However, we discuss the effective exponents as functions of
the flow parameter.  Then the effective exponents are simply  given by
the expressions for the asymptotic exponents (\ref{10}) but with
replacing the fixed point values of the couplings $u^*$ and $v^*$ by
the solutions of the flow equations (\ref{6}):
\begin{equation}
1/\nu_{eff} = 2(1-\gamma_m(u(\ell), v(\ell))),\,\,\,\,\,
 \eta_{eff} = 2\gamma_2(u(\ell), v(\ell)),\,\,\,\,\,
 \gamma_{eff}=\nu_{eff}(2-\eta_{eff})
\end{equation}

\begin{figure}[htbp]
\begin{centering}
\setlength{\unitlength}{1mm}
\begin{picture}(150,120)
\epsfxsize=150mm
\epsfysize=120mm
%\put(-8,125)
{\epsffile[50 50 750 554]{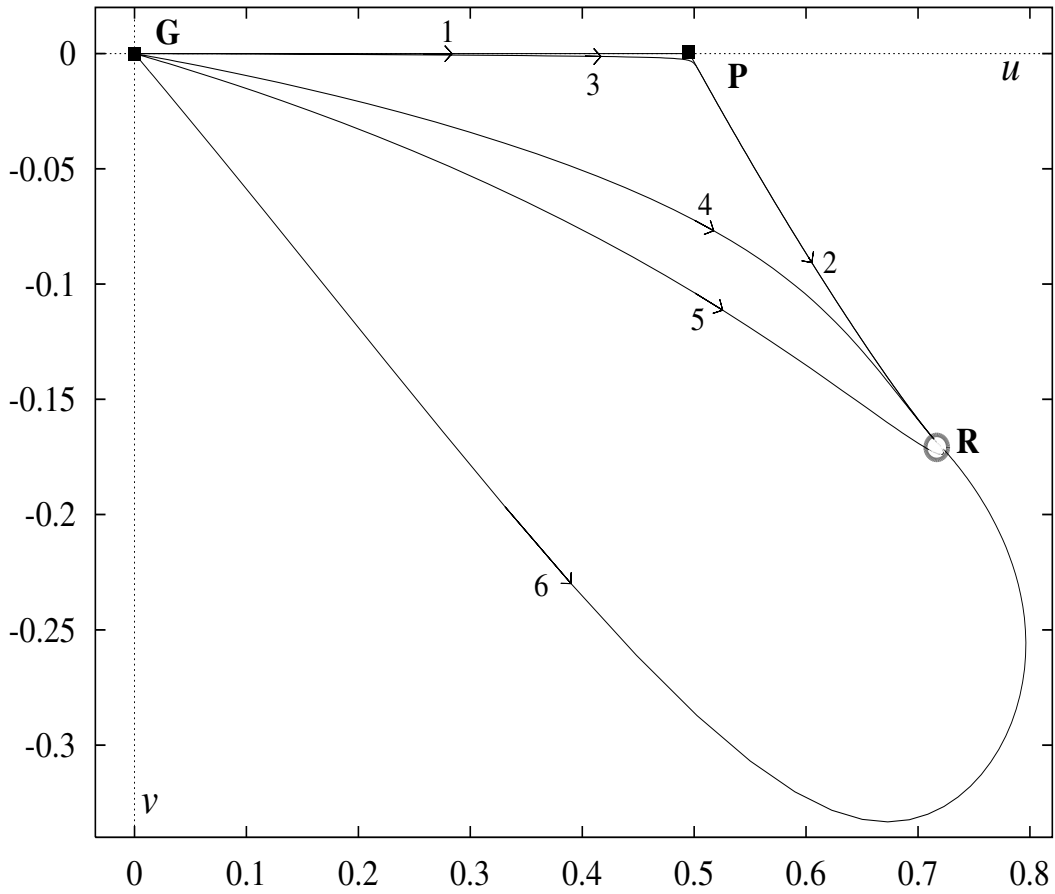}}
\end{picture}
\end{centering}
\caption{\protect\label{flows.eps}
{
Flow lines for the $d=3$ RIM. Fixed points $G, P$ are unstable,
fixed point $R$ is the stable one.
}}
\end{figure}

\begin{figure}[htbp]
\begin{centering}
\setlength{\unitlength}{1mm}
\begin{picture}(150,120)
\epsfxsize=150mm
\epsfysize=120mm
%\put(-8,125)
{\epsffile[50 50 750 554]{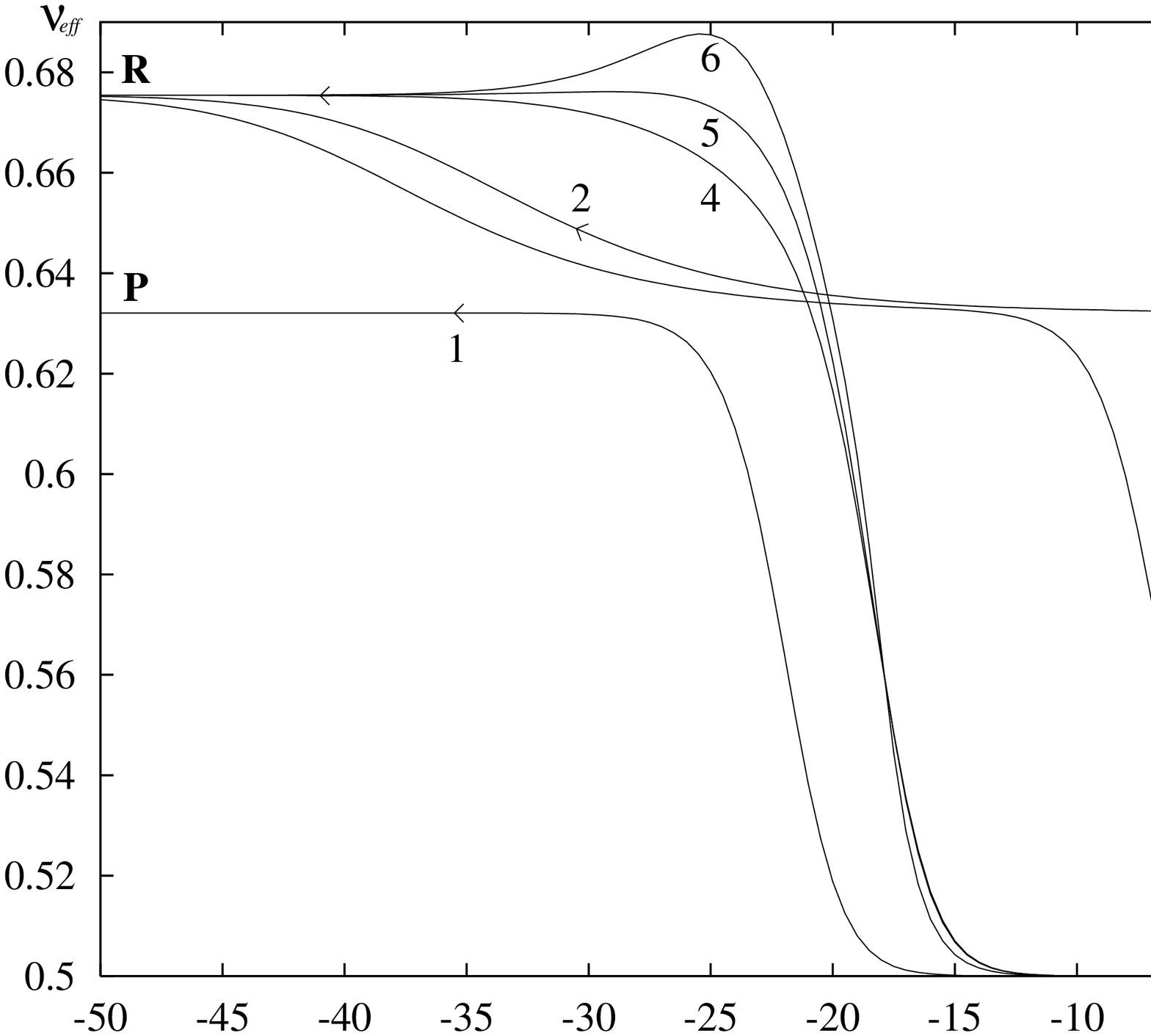}}
\end{picture}
\end{centering}
\caption{\protect\label{nu.eps}
{
Effective exponent $\nu_{eff}$ versus logarithm of the flow parameter
$\ell$ for the flows shown in Fig. \protect\ref{flows.eps}.
}} \end{figure}

\begin{figure}[htbp]
\begin{centering}
\setlength{\unitlength}{1mm}
\begin{picture}(150,120)
\epsfxsize=150mm
\epsfysize=120mm
%\put(-8,125)
{\epsffile[50 50 750 554]{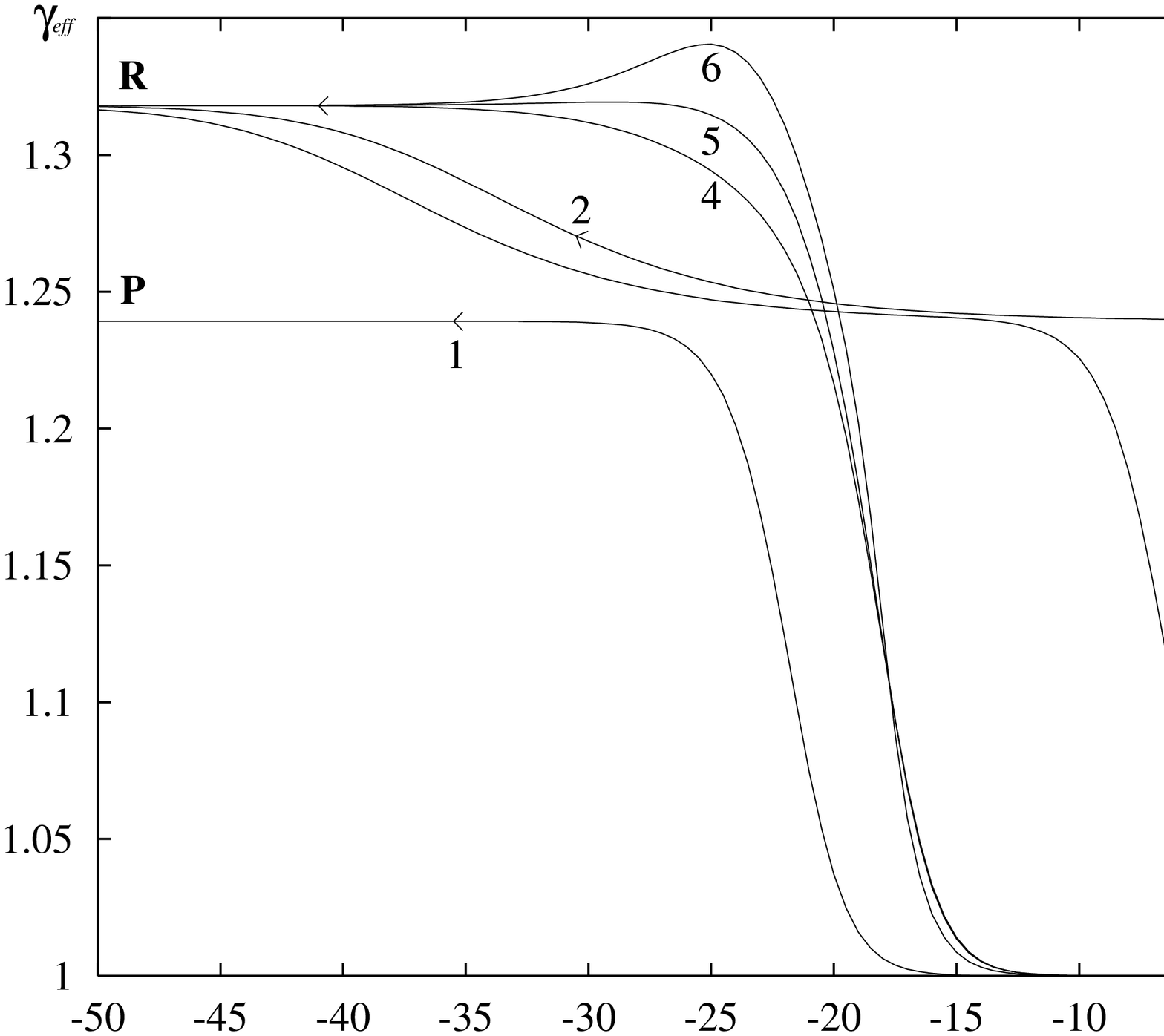}}
\end{picture}
\end{centering}
\caption{\protect\label{gamma.eps}
{
Effective exponent $\gamma_{eff}$ versus logarithm of the flow
parameter $\ell$ for the flows shown in Fig. \protect\ref{flows.eps}.
}}
\end{figure}

These solutions are shown in  Fig. \ref{flows.eps} for several initial
conditions. There are shown the two unstable fixed points:  G
(Gaussian fixed point) and P (pure Ising fixed point) and the stable
random fixed point R, with both couplings non zero. In the background
region the couplings are expected to be small thus all the flows shown
start near the Gaussian fixed point with different ratios
$v(\ell_0)/u(\ell_0)$ (curves 1 to 6 except curve 2). Curve 1 is the
separatrix connecting the fixed point G with P, and curve 2 is the
separatrix connecting P with R.  All curves starting with a negative
coupling $v$ remain negative but the dependence might be nonmonotonous
(see curve 6). Thus several scenarios for the values of the effective
critical exponents are possible (see Figs. \ref{nu.eps},
\ref{gamma.eps}).

Both in experiment as well as in computer simulations (see Tabs.
\ref{table_experiment}, \ref{table_MC}) exponents reported differ and
even exceed the known asymptotic theoretical values. This nonuniversal
behavior might be related to the possible nonasymptotic behavior found
in our different flows as has been suggested in
\cite{Heuer93,Janssen95}.  The difference might be due to (i) the
different temperature regions of the experiment and/or (ii) the
different concentrations. The initial values for the couplings in the
flow equations depend on the value of the concentration, especially
for small dilution one expects the coupling $v$ to be proportional to
the concentration. If this is the case we expect a monotonous increase
of the values of the effective exponent to the asymptotic value. A
typical scenario is seen in curve 3 of Figs.  \ref{flows.eps},
\ref{nu.eps}, \ref{gamma.eps}.  In this case effective exponents equal
to the pure model critical exponents might be found in relatively wide
region of temperature. Then as the attraction region of the fixed
point P becomes weaker and weaker, an overshooting is possible and
effective exponents larger its asymptotic values might be found.  This
scenario is predicted for larger dilutions and represented by the
curve 6 of Figs.  \ref{flows.eps}, \ref{nu.eps}, \ref{gamma.eps}.
Curves 5, 6 correspond to situation when crossover from the mean field
behavior towards the random one is not influenced by the presence of a
pure fixed point.

%%%%%%%%%%%%%%%%%%%%%%%%%%%%%%%%%%%%%%%%%%%%%%%%%%%%%%%%%%%%%%%%%%%%
%                              SECTION VI
%%%%%%%%%%%%%%%%%%%%%%%%%%%%%%%%%%%%%%%%%%%%%%%%%%%%%%%%%%%%%%%%%%%%
\section{Conclusions \label{VI}}

We studied the critical behavior of the three-dimensional weakly
diluted Ising model asymptotically close to the critical point, in the intermediate region
and far from it. To this end we calculated the
values of the asymptotical critical exponents, analyzed the behavior
of the effective critical exponents and obtained the value of the
correction-to-scaling exponent entering the Wegner expression which
describes the approach of singular quantities to the critical
temperature. Our study is based on the 5--loop minimal subtraction
field-theoretical renormalization group functions \cite{Kleinert95} of
the $\phi^4$--theory with $O(n)$--symmetric and cubic interactions
which in the replica limit $n \rightarrow 0$ correspond to the diluted
Ising model case. As the minimal subtraction scheme allows to develop
either the $\sqrt{\varepsilon}$--expansion or the 3d approach, we
adopted the both schemes and compared the results obtained on their basis.

The perturbation theory RG functions series are asymptotic at best.
In order to calculate the critical exponents we adopt different
resummation procedures within the both schemes, testing  them
on the well established case of the $\phi^4$--model with one
coupling (pure Ising model). While after the resummation both
$\varepsilon$--expansion and 3d approach give reliable results for the
numerical values of the pure model critical exponents, in the case of
the diluted Ising model the bad convergence properties of the
$\sqrt{\varepsilon}$--expansion do not allow to obtain reliable
values at $d=3$. Using the direct resummation of the RG functions at
$d=3$ in the minimal subtraction scheme we obtained numerical values
for the asymptotic critical exponents of the diluted Ising model and
estimated their accuracy to be of order of several percents. The results
obtained in the 3d approach  agree well with other theoretical
and recent Monte Carlo simulations.  Studying the crossover effect we
calculated the effective critical exponents and their flows in
the nonasymptotic region. We observed several scenarios of crossover
in the RIM including: monotonous crossover from the mean field values
of critical exponents to the random ones; existence of a wide
temperature region where the RIM  effective exponents coincide with
the asymptotic exponents of the pure Ising model; possible values of
effective exponents exceeding those of asymptotic ones.

Though the 3d approach of calculation encountered difficulties in the
5--loop level we guess that the fixed $d$ approach, both within the
massive \cite{Parisi80} and minimal subtraction
\cite{tHooft72,Schloms} schemes, remains the only reliable way to
study critical behavior of the model by means of the RG technique.

{\it Note added in proof.}
During processing the article several new results appeared in
the field. The massive scheme 4--loop RG functions of the RIM
\cite{Mayer89,Mayer89a}
were extended to the five-loop
\cite{Pakhnin99}
and to the six-loop level
\cite{Carmona99}. While the traditional analysis of the first via PB resummation
allowed to obtain numerical values for the RIM asymptotic critical
exponents
\cite{Pakhnin99}, the method failed for the higher-order functions
\cite{Folk_unp}.
However, an application to the functions of a refined resummation
procedure which treats renormalized couplings of the RG functions
asymmetrically
\cite{Alvarez00}
reconstituted and improved
\cite{Pelissetto00}
earliear data for
the RIM asymptotic critical exponents.

\begin{center}
{\bf Acknowledgements}
\end{center}
We acknowledge helpful discussions with  Hagen Kleinert and
Verena Schulte-Froh\-lin\-de.
This work has been supported in part by "\"Osterreichische Nationalbank
Ju\-bi\-l\"a\-ums\-fonds" through the grant No 7694.

%%%%%%%%%%%%%%%%%%%%%%%%%%%%%%%%%%%%%%%%%%%%%%%%%%%%%%%%%%%%%%%%%%%%
%                              APPENDIX A
%%%%%%%%%%%%%%%%%%%%%%%%%%%%%%%%%%%%%%%%%%%%%%%%%%%%%%%%%%%%%%%%%%%%

\section{Appendix A: The Choice of the Resummation Procedure
}
In order to obtain reliable quantitative description of the problem
under consideration one should adjust precisely resummation procedures
necessary for analysis of RG functions.  Since no information is
available about the high order behavior of the series for $\beta$--
and $\gamma$--functions (\ref {11})-(\ref{14}) (compare with the $d=2$
and $d=3$ scalar $\phi^4$--model where the Borel summability of the RG
functions is proven \cite{borelsummability,note} and the large order
behavior is calculated \cite{Lipatov77,Brezin77}), we reject all
powerful methods implementing such an information (e. g. resummation
refined by a conformal mapping, widely used in the models of critical
phenomena with one coupling \cite{Guillou80, Guida98})
and restrict ourselves to the simplest resummation techniques which
are the generalisation of the Pad\'e--Borel technique
(\ref{18})--(\ref{Pade-Borel}) to the two variable case. Among them we
choose that procedure, which, for the different orders of approximation, provides
the maximal stability. That means, the fastest convergence of the results as well
as the maximal similarity of the topological structure of the lines defined by the
zeros of the $\beta$--functions.

\begin{center}
a) A $d=0$ theory
\end{center}
Let us start tuning the resummation technique by considering the
expressions for a toy--model. It is defined by the partition function
$$
Z=\frac{1}{\pi^{n/2}} \int_{-\infty}^{+\infty} {\rm d}
\phi_1\ldots{\rm d} \phi_n
\exp(-\sum_{i=1}^n{\phi_i^2}-u(\sum_{i=1}^n{\phi_i^2})^2
-v\sum_{i=1}^n{\phi_i^4}) \, ,
$$
which corresponds to the cubic model described by the Hamiltonian (\ref{1})
in dimension $d=0$. One can easily
calculate the sufficient number of terms representing $Z$ in the form
of a series in $u$ and $v$ for arbitrary $n$. However, the simplest
case which reproduces the series in two couplings is $n=2$ since the
case $n=0$ is trivial and $n=1$ corresponds to a series in a single
variable $u+v$.

\begin{figure}[htbp]
\begin{centering}
\setlength{\unitlength}{1mm}
\begin{picture}(140,100)
\epsfxsize=120mm
\epsfysize=100mm
\put(0,5){\epsffile[50 50 554 770]{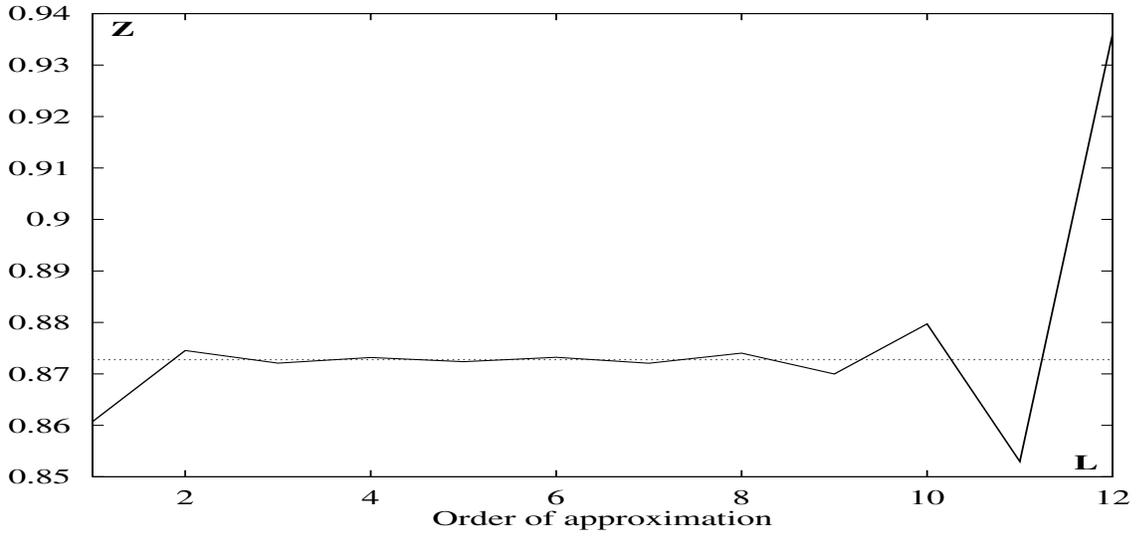}}
\end{picture}
\end{centering}
%\vspace{-5mm}
\caption{\label{lin.eps}
The estimate of the toy--model partition function $Z$ at $u=1/10$ and
$v=-1/100$ in dependence of the order of perturbation
theory $L$ in couplings $u$ and $v$. The application of Pad\'e--Borel
resummation with linear denominator approximants of type $\left[
N/1\right]$  (solid line) provides convergence to the exact value
(dotted line) only for some first orders of approximation.}
\end{figure}

\begin{figure}[htbp]
\begin{centering}
\setlength{\unitlength}{1mm}
\begin{picture}(140,100)
\epsfxsize=120mm
\epsfysize=100mm
\put(0,5){\epsffile[50 50 554 770]{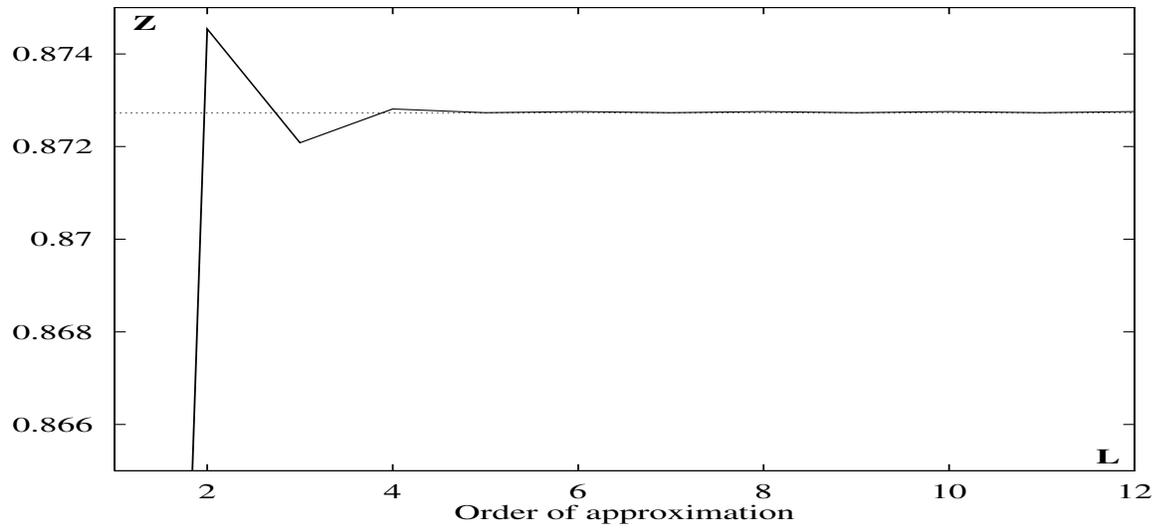}}
\end{picture}
\end{centering}
%\vspace{-5mm}
\caption{\label{diag.eps}
The estimate of the toy-model partition function $Z$ at $u=1/10$ and
$v=-1/100$ in dependence of the order of perturbation
theory $L$ in couplings $u$ and $v$.  The application of Pad\'e--Borel
resummation with diagonal approximants of type $\left[ N/N\right]$
or $\left[ (N+1)/N\right]$ (solid line) provides perfect convergence
of the estimate to the exact value (dotted line).}
\end{figure}

We write the series for $Z$ in the two variables $u$ and $v$ as a series in
one auxiliary variable $\tau$ introduced by the substitution $u
\rightarrow u\tau$, $v \rightarrow v\tau$ (the so called resolvent series
\cite{Watson74}) and apply the Pad\'e--Borel (PB) method
(\ref{18})--(\ref{Pade-Borel}) to the series in $\tau$. Here we choose
two possibilities of the approximants, one with linear denominator and
the other one of diagonal type.
Moreover we fix the parameter $p$ (\ref{18}) to $p=0$ but discuss other
choices for the parameter later on. The results of this procedure are shown in
Figs.  \ref{lin.eps}---\ref{diag.eps} for the specific values $u=0.1$
and $v=-0.01$.  One notes that a stable convergence to the exact
value takes place only for resummations with diagonal approximants, while
the approximants with linear denominator converge only for certain,
``optimally truncated'', orders of approximation.
The larger the value of the variables $u$ and $v$ the
less is the order of ``optimal truncation''. We believe that
processing the divergent series (\ref{11})--(\ref{14}) in the same manner
one may encounter a similar difficulty.

Analysing the toy--model series by the Chisholm-Borel (CB) scheme
(\ref{20})--(\ref{21}), one notes that the coefficients in the Chisholm
approximant are underdetermined. For example, for the $L$--loop sum an
approximant of the linear--denominator type $\left[ L/1 \right]$ beeing
uniquely defined requires two additional equations. This is the minimal
number of additional conditions, necessary, to determine any approximant for
$L=2 \ldots 5$ except for $L=4$, where the $\left[ 3/2 \right]$ approximant is
determined uniquely without additional conditions.  Since, generally speaking,
a near--diagonal
Chisholm approximant requires additional conditions the number of which
depends  on the order of approximation, we reject this type of approximants
and consider in the following only the $\left[ L/1 \right]$ type approximants.
The two additional equations have to be symmetric in the variables $u$ and $v$,
otherwise the properties of the symmetry related to
these variables would depend on the method of calculation.
By the substitution $v=0$ all
the equations describing the critical behavior of the diluted model
are converted into the appropriate equations of the pure model. However,
if a pure model is solved independently, the resummation technique
uses the Pad\'e approximant. Thus, the Chisholm
approximant has to be chosen in such a way that, for either $u=0$
or $v=0$ one obtains the Pad\'e approximant of the
one-variable case. This also needs a special choice of the additional
conditions. This is achieved by choosing  the Chisholm
approximant $\left[ L/1 \right]$ with the numerator coefficients
at $u^L$ and $v^L$ equal to zero.

The analysis of the toy--model series by means of this type of CB
technique leads to the existence of an ``optimally truncated'' order of
approximation similarly to the one shown in Fig. (\ref{diag.eps}).

\begin{center}
b) A $d=3$ theory
\end{center}
Let us turn back now to the expressions for the RG functions
(\ref{11})-(\ref{14}) at $d=3$. Starting from the PB analysis one
finds that diagonal--type approximants lead to poles on the real axis.
This may be due to the fact that such approximants do not
reconstitute the sign-alternating high-order behavior of the general
term of the RG functions, which was confirmed in the particular cases
$n=2$ and $n=3$ \cite{Kleinert95a}.  Thus for the further discussions we
proceed with approximants of linear--denominator type.

Applying the PB method with linear--type approximants we find that
the random fixed point for the $d=3$ RIM cannot be reconstituted
already in the 3--loop approximation, while the  random fixed point
exists even for nonresummed $\beta$--functions in this approximation. The
picture does not change qualitatively when we try to increase
effectively the order of polynomial representation for the
$\beta$--functions by resumming expressions $[1+r_u(u+v)]\beta_u,
[1+r_v(u+v)]\beta_v$ with $r_u, r_v$ taken as fit parameters (compare with
\cite{Guida98}).  The modified construction of Ref. \cite{Janssen95}
for the Pad\'e approximant with a linear denominator
reconstituting the known large order behavior of the one coupling
$\beta$--functions, $\beta_u(u,v=0)$ and $\beta_v(u=0,v)$,  does not
lead to the appearance of the random fixed point either. We conclude,
although the PB scheme works in 2-- and 4--loop
approximations, it appears to be unstable in the 3--loop approximation
and therefore we eliminate it from our consideration.

The CB method reconstitutes the random fixed point in 2--, 3-- and
4--loop approximation, however it fails in the 5--loop approximation. In
order to reestablish the random fixed point we have varied the values of $p$
as well as of $r_u, r_v$, but even then we find no region for
$r_u>0, r_v>0$ so that the random fixed point exists in all orders
of perturbation theory.
We also predicted the values of the sixth order contribution for
$\beta$--functions (see Refs. \cite{Samuel93,Samuel95}), but the
resummation of such  pseudo--6--loop functions did  not
allow to find the random fixed point either. Again a modified construction
of the Chisholm approximant with the known linear denominator
\cite{Janssen95} failed. Comparing the behavior of the
toy--model series (see Fig. \ref{lin.eps}) and the convergence of the
results of the RIM (Tables \ref{tab8}) we consider
$L=4$ as the ``optimal truncation'' order within the CB scheme
of linear--denominator approximant.

\begin{figure}[htbp]
\begin{centering}
\setlength{\unitlength}{1mm}
\begin{picture}(140,100)
\epsfxsize=120mm
\epsfysize=100mm
\put(0,5){\epsffile[50 50 554 770]{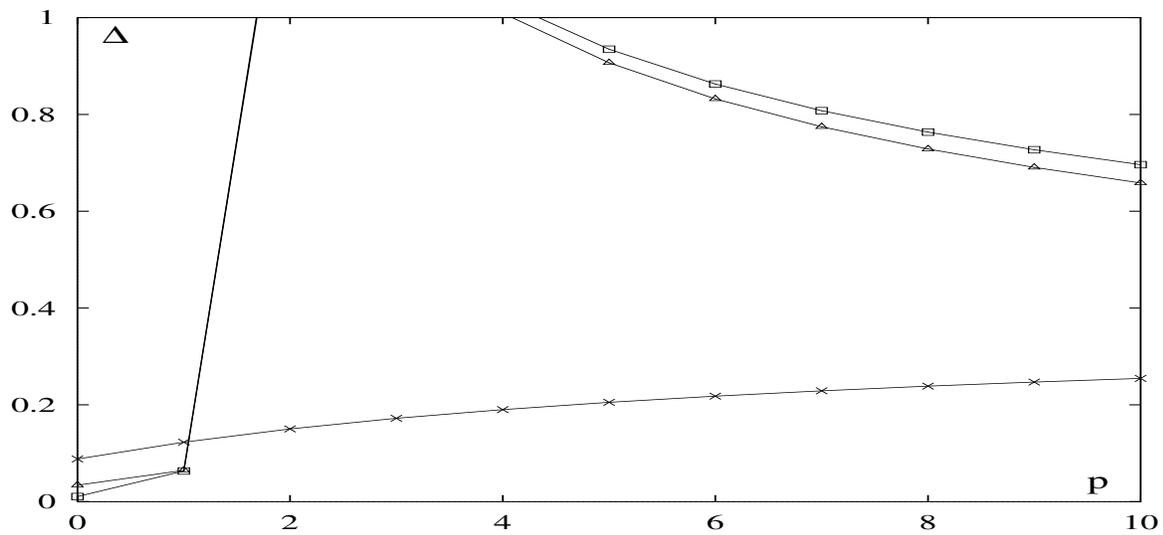}}
\end{picture}
\end{centering}
%\vspace{-5mm}
\caption{\label{opt.eps}
The dependence of a measure $\Delta$ of total deviation between $3$--
and $4$--loop pure Ising model results (triangles) as well as of $3$--
and $4$--loop RIM results (crosses) on fitting parameter $p$. One can
see the minimum for $p=0$. The deviation between $5$-- and $4$--loop
results (boxes) for pure Ising model confirms $p=0$, too.  }
\end{figure}

Once we have chosen the CB method based on Chisholm approximants of $\left[
L/1 \right]$ type as the tool for analysing the RIM RG functions we
look now for the fastest convergence of the results with increase of
the order of approximation in number of loops $L$. To this end we fit
the parameters $p$ entering Borel-Leroy images (\ref{20}) of RG
functions.  For the RIM we introduce a measure of total deviation
between $L$-- and $L'$--loop results by a function
$\Delta^2=(u^{*,L}-u^{*,L'})^2+(v^{*,L}-v^{*,L'})^2+
(\gamma^{*,L}-\gamma^{*,L'})^2+(\nu^{*,L}-\nu^{*,L'})^2$, where the
superscripts $L(L')$ indicate a value obtained in $L(L')$--loop
approximation. For the pure Ising model an appropriate measure is given by
a similar function with $v^*=0$. We adjust now the parameter $p$ to
minimize $\Delta$. The behavior of $\Delta(p)$ is shown in
Fig. \ref{opt.eps}.  A minimum of $\Delta$ for the pure Ising model is
achieved for $p=0$ (curves pointed by boxes and triangles in Fig.
\ref{opt.eps}) in both cases $L=4, L'=3$ and $L=5, L'=4$. Similar behavior
in $p$ is observed for the RIM (curve pointed by crosses) suggesting
the choice $p=0$ as well.

\begin{figure}[htbp]
\begin{centering}
\setlength{\unitlength}{1mm}
\begin{picture}(126,200)
\epsfxsize=126mm
\epsfysize=200mm
\put(0,10){\epsffile[23 10 567 835]{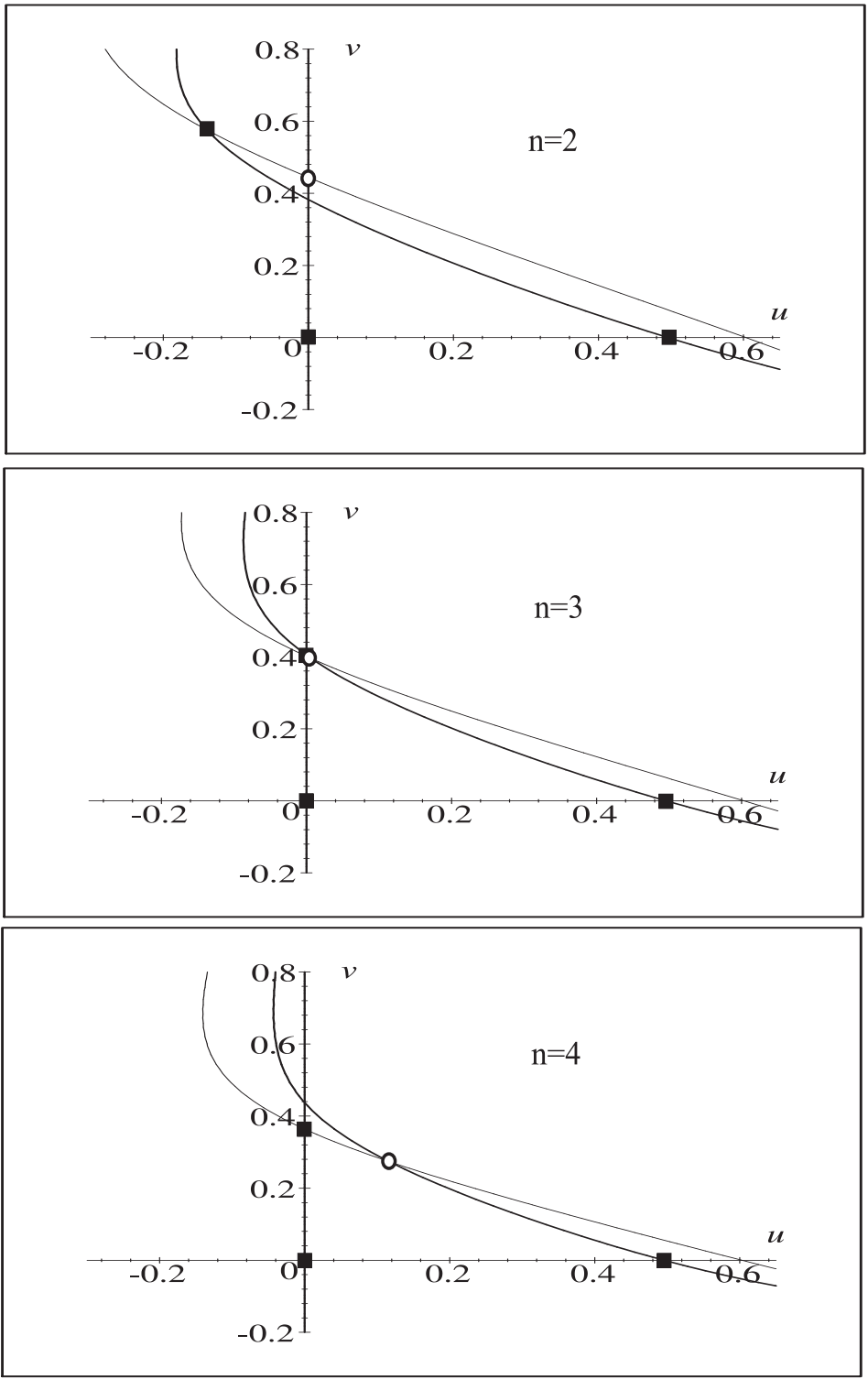}}
\end{picture}\\
\end{centering}
\caption{\label{cubic}
The lines of zeros of the cubic model $\beta$--functions resummed by
the Chi\-sholm-Borel method in 4--loop approximation for different $n$.
Thick lines correspond to $\beta_u=0$, thin lines depict $\beta_v=0$.
The filled boxes and open circles show the positions of unstable and
stable fixed points respectively. One can see that the crossover to a
new behavior appears for values of $n$ very close to $3$. Our estimate
yields $n_c=2.950$ (see the text).}
\end{figure}

\begin{center}
c) A cubic model
\end{center}
Since originally the RG functions
under consideration (\ref{11})--(\ref{14}) were obtained in order to
study the critical behavior of the cubic model introduced by the
effective Hamiltonian (\ref{1}), we will use it for another test of
the resummation procedure.  In the case of nonzero values of $n$ the
critical behavior of this system is governed either by the
$O(n)$--symmetric fixed point for  values of $n$ small enough or
by the cubic
fixed point for $n>n_c$, where $n_c$ is the marginal value
of the order parameter component. The identification of the
marginal dimension $n_c$ as well as of the critical exponents
governing the phase transition of the model was recently performed
within the $\varepsilon$--expansion \cite{Kleinert95}.
The same model was studied by means of the massive RG approach
\cite{Mayer89}. The numerical value of $n_c$ was found to be only
slightly less than $n=3$ leading to values of the critical exponents of
the cubic model practically indistinguishable from those of the
$O(3)$--symmetric model.  Recent MC simulations strongly
suggest $n_c=3$ although $n_c<3$ cannot be excluded \cite{Caselle98}.  We
present estimations of $n_c$ obtained from the conditions that
the $O(n)$--symmetric and the cubic fixed point  of the
resummed $\beta$--functions coincide. This calculation has two
advantages: (i) one tests once more the resummation methods and, (ii)
a new estimate of $n_c$ is obtained from the 3d approach within
minimal subtraction scheme. Up to now $n_c$ has been calculated in the
frames of $\varepsilon$--expansion in the 5--loop approximation
\cite{Kleinert95,Shalaev97} as well as in the massive scheme in 4--loop
approximation
\cite{Mayer89}. We perform  estimates within the 3d minimal subtraction
approach \cite{Schloms} in 2--, 3-- and 4--loop approximation. The
corresponding values  on the basis of $\beta$--functions resummed by the
Chisholm-Borel method read:  $n_c(2Loops)=2.7730,
n_c(3Loops)=3.1078, n_c(4Loops)=2.9500$.  This should be compared with
the result of the 5--loop $\varepsilon$--expansion $n_c=2.958$ (Pad\'e
analysis) \cite{Kleinert95}, $n_c=2.855$ (Pad\'e-Borel resummation)
\cite{Shalaev97} and the 4--loop massive 3d RG scheme $n_c=2.90$
(Chisholm-Borel resummation) \cite{Mayer89}. In Fig.  \ref{cubic}
we show the lines of zeros for the resummed $\beta-$functions as well
as the fixed points for different $n$. One
can see that the change of stability of the fixed points
appears for $n$ very close to $n=3$.  As
our numerical result yields $n_c<3$ the cubic model at $n=3$ is
governed by a {\bf new set} of critical exponents which read
$\gamma=1.387, \nu=0.709, \alpha=-0.127, \eta=0.044$. The coordinates
of the stable cubic fixed point  $u^*=0.0064, v^*=0.3950$ and stability
matrix eigenvalues $\omega_1=0.0440, \omega_2=0.0055$ should be
compared with the corresponding values for the nearby situated unstable
$O(n)$--symmetric fixed point:  $u^*=0, v^*=0.4009, \omega_1=-0.0056,
\omega_2=0.0751$.  Of course no experiment can distinguish between the
critical exponents of these almost coinciding fixed points.
Note however that at $n>n_c$ the stable fixed
point can be reached only for $u>0$. Then $n_c<3$ means that all
systems described by the phenomenological Landau-Ginsburg Hamiltonian with
cubic anisotropy of negative coupling should undergo a weak first-order phase
transition at $n=3$ \cite{Sznajd}.

%%%%%%%%%%%%%%%%%%%%%%%%%%%%%%%%%%%%%%%%%%%%%%%%%%%%%%%%%%%%%%%%%%%%
%                              THE BIBLIOGRAPHY
%%%%%%%%%%%%%%%%%%%%%%%%%%%%%%%%%%%%%%%%%%%%%%%%%%%%%%%%%%%%%%%%%%%%

%%%%%%%%%%%%%%%%%%%%%%%%%%%%%%%%%%%%%%%%%%%%%%%%%%%%%%%%%%%%%%%%%%%%
%                              TABLES
%%%%%%%%%%%%%%%%%%%%%%%%%%%%%%%%%%%%%%%%%%%%%%%%%%%%%%%%%%%%%%%%%%%%
\newpage
{\footnotesize
\begin{table}
\caption{\label{table_experiment}
{The experimentally measured critical exponents of the materials,
which correspond to the $d=3$ RIM. The measurement procedures as well
as materials specifications are given in the following notations:
NMR --- nuclear magnetic resonance;
LB --- linear birefringence;
NS --- neutron scattering;
MS --- M\"ossbauer spectroscopy;
SMXS --- synchrotron magnetic x-ray scattering;
XS --- x-ray scattering;
DAG --- dysprosium aluminium garnet;
$\tau_{min}$ denotes the minimal value of the reduced temperature
reached in an experiment.
}
}
% the structure of the Table:
% Ref, year, material, method, minimal tau, beta, gamma, nu, alpha
\begin{center}
\tabcolsep0.6mm % Note - you may change separator between column
\begin{tabular}{|c|c|c|c|c|c|c|c|c|}
\hline
\hline
Ref. & year & material & method &$\tau_{min}$& $\beta$ & $\gamma$&
$\nu$ & $\alpha$ \\
\hline
\cite{Dunlap81}&
       1981&
              $Mn_xZn_{1-x}F_2,$&
                          NMR&
                                     $?$&
                                        0.349$\pm$0.008
                                             & --- & ---& ---
\\
&&            $x=0.864$                       &&&&&&
\\
\hline
\cite{Birgeneau83}&
       1983&
               $Fe_xZn_{1-x}F_2,$&
                         NS; LB&
                                  $2 \cdot 10^{-3}$&
                                   --- &
                                            1.44$\pm$0.06&
                                                       0.73$\pm$0.03&
                                                       $-0.09\pm0.03$
\\ &&           $x=0.6, 0.5$                      &&&&&& \\
\hline
\cite{Hastings85}&         1985& DAG& NS& $4 \cdot 10^{-2}$&
0.350$\pm$0.01& --- & 0.73 & --- \\ &&             +1\% Y powder
&&&&&& \\ \hline \cite{Belanger86}& 1986&
               $Fe_xZn_{1-x}F_2,$& NS& $1.5 \cdot 10^{-3}$& --- &
                                   1.31$\pm$0.03& 0.69$\pm$0.01& \\ &&
$x=0.46$ &&&&&& \\ \hline
\cite{Barret86}& 1986& $Fe_xZn_{1-x}F_2,$& MS&
                              $10^{-3}$& 0.36$\pm$0.01& --- & ---& ---
\\ &&               $x=0.9925-0.95$ &&&&&&
\\ \hline
\cite{Mitchell86}& 1986&
                $Mn_xZn_{1-x}F_2,$& NS&
                              $4 \cdot 10^{-4}$& --- &
                                     1.364$\pm$0.076& 0.715$\pm$0.035&
                                                     --- \\
&&                $x=0.75$ &&&&&& \\
\hline \cite{Thurston88}&
             1988& $Mn_xZn_{1-x}F_2,$&
                               SMXS& $10^{-3}$&
                                      0.33$\pm$0.02& --- & ---& ---
\\ &&                 $x=0.5$ &&&&&&
\\ \hline
\cite{Rosov88}& 1988&
                 $Fe_xZn_{1-x}F_2,$& MS&
                               $3 \cdot 10^{-4}$& 0.350$\pm$0.009&
                                                       --- & ---& ---
\\
&&                $x=0.9$ &&&&&&
\\
\hline
\cite{Ramos88} &
            1988&
                 $Mn_xZn_{1-x}F_2,$&
                              LB&  $< 10^{-2}$ &   --- &
                                 --- & ---& -0.09$\pm$0.03
\\
&&                $x=0.40, 0.55, 0.83$ &&&&&&
 \\
\hline
\cite{Ferreira91} &
            1991&
                 $Fe_xZn_{1-x}F_2,$&
                              LB&  ? &   --- &
                                 --- & ---& -0.09
\\
&&                $0.31\leq x \leq 0.84$ &&&&&&
  \\
\hline
\cite{Belanger95} &
            1995&
                 $Fe_xZn_{1-x}F_2,$&
                              NS&  $<10^{-1}$& 0.35 &
                                 --- & ---& ---
\\
&&                $x =0.5$ &&&&&&
  \\
\hline
\cite{Belanger96} &
            1996&
                 $Fe_xZn_{1-x}F_2,$&
                              NS&  $10^{-2}$&  0.35 &
                                 --- & ---& ---
\\
&&                $x =0.52$ &&&&&&
  \\
\hline
\cite{Hill97} &
            1997&
                 $Fe_xZn_{1-x}F_2,$&
                              XS&  $10^{-2}$&  0.36$\pm$ 0.02&
                                 --- & ---& ---
\\
&&                $x =0.5$ &&&&&&
\\
\hline
\cite{Slanic98a} &
            1998&
                 $Fe_xZn_{1-x}F_2,$&
                              LB&  $2.5\cdot 10^{-3}$& --- &
                                 --- & ---& $-0.10\pm0.02$
\\
&&                $x =0.93$ &&&&&&
  \\
\hline
\cite{Slanic98} &
            1998&
                 $Fe_xZn_{1-x}F_2,$&
                              NS&  $2\cdot 10^{-3}$&  ---&
                                 1.35$\pm$0.01 & 0.71$\pm$0.01& ---
\\
&&                $x =0.93$ &&&&&&
  \\
\hline
\cite{Slanic99} &
            1999&
                 $Fe_xZn_{1-x}F_2,$&
                              NS&  $1.14\cdot 10^{-4}$&  ---&
                                 1.34$\pm$0.06 & 0.70$\pm$0.02& ---
\\
&&                $x =0.93$ &&&&&&
\\
\hline \hline
\end{tabular}
\end{center}
\end{table}
}

{\footnotesize
\begin{table}
\caption{\label{table_MC}
{The critical exponents of the $d=3$ RIM as they are
obtained in $MC$ simulations for different values of magnetic sites
concentration $x$.  Maximal number of lattice sites simulated is
$N=L^3$.  (The asterisk at concentrations denotes that disorder was
realized in a grand-canonical manner.)} }
\begin{center}
\tabcolsep0.6mm
\begin{tabular}{|c|c|c|c|c|c|c|}
\hline
\hline
Ref.& year & $L$ & $x$ & $\beta$ & $\gamma$ & $\nu$
\\
\hline
\cite{Landau80} & 1980 &  30  &  $0.4 < x \leq 1$ &  $0.31$ & $1.25$
&
\\
\hline
                &      &      & 1  & $0.30 \pm 0.02$ &---&---
\\
                &      &      & 0.985  & $0.31 \pm 0.02$ &---&---
\\
\cite{Marro86}  & 1986 &  40  & 0.95   & $0.32 \pm 0.03$ &---&---
\\
                &      &      & 0.9    & $0.355 \pm 0.010$ &---&---
\\
                &      &      & 0.8    & $0.385 \pm 0.015$ &---&---
\\
\hline
                &      &       & 1     & $0.29 \pm 0.02$ &---&---
\\
\cite{Chowdhury86}& 1986 & 90  & 0.95   & $0.28 \pm 0.02$ &---&---
\\
                &      &       & 0.90   & $0.31 \pm 0.02$ &---&---
\\
                &      &       & 0.80   & $0.37 \pm 0.02$ &---&---
\\
\hline
\cite{Braun88}  & 1988&   40   & 0.80   & $0.392 \pm 0.03$ &---&---
\\
\hline
\cite{Wang89}   & 1989 &  100  & $0.4 \leq x \leq 0.8$   &---&
 $1.52 \pm 0.07$ & $0.77 \pm 0.04$
\\
\hline
\cite{Wang90}   & 1990 & 300   & 0.8   &---&$1.36\pm 0.04$&---
\\
\hline
                &      &       & 1     & --- & --- & 0.629(4)
\\
\cite{Holey90}  & 1990 &  64   & 0.9   & --- & --- & $ < 2/3$
\\
                &      &       & 0.8   & --- & --- & 0.688(13)
\\
\hline
                &      &       & 1
& $0.305\pm0.01$ & $1.24\pm0.01$&---
\\
                &      &       & 0.9   & $0.315\pm0.01$
& $1.30\pm0.01$&---
\\
\cite{Heuer90}  & 1990 &  60   & 0.8   & $0.330\pm0.01$
& $1.35\pm0.01$&---
\\
                &      &       & 0.6   & $0.330\pm0.01$
& $1.48\pm0.02$&---
\\
                &      &       & 0.5   & $0.335\pm0.01$
& $1.49\pm0.02$&---
\\
\hline
                &      &       & 1     & $0.33\pm 0.01$
& $1.22\pm0.02$ & $0.624 \pm 0.010$
\\
                &      &       & 0.95  & $0.31\pm0.02$
& $1.28\pm0.03$ & $0.64 \pm 0.02$
\\
\cite{Heuer93}  & 1993 &  60   & 0.9   & $0.31\pm0.02$
& $1.31\pm0.03$ & $0.65 \pm 0.02$
\\
                &      &       & 0.8   & $0.35\pm0.02$
& $1.35\pm0.03$ & $0.68 \pm 0.02$
\\
                &      &       & 0.6   & $0.33\pm0.02$
& $1.51\pm0.03$ & $0.72 \pm 0.02$
\\
\hline
\cite{Hennecke93}& 1993 &  90? & 0.6 & $0.42 \pm 0.04$
& --- &$0.78 \pm 0.01$
\\
\hline
\cite{Wiseman98a} & 1998 & 64  & 0.8 & $0.344 \pm 0.003$
& $1.357 \pm 0.008$ & $0.682 \pm 0.003$
\\
\hline
\cite{Wiseman98b} & 1998 & 90  &  0.6 & $0.316 \pm 0.013$
& $1.522 \pm 0.031$ & $0.722 \pm 0.008$
\\
                  &      & 80  & 0.6$^*$ & $0.313 \pm 0.012$
& $1.508 \pm 0.028$ & $0.717 \pm 0.007$
\\
\hline
\cite{Parisi98} & 1998 & 128 & $0.4 \leq x \leq 0.9$ & $0.3546
\pm 0.0028$
& $1.342 \pm 0.010$ & $0.6837 \pm 0.0053$
\\
\hline
\end{tabular}
\end{center}
\end{table}
}

\begin{table}
\begin{center}
\tabcolsep1.6mm
\begin{tabular}{||c|cccccc||}
\hline
${\bf \nu}$ & 0 & 1 & 2 & 3 & 4 & 5 \\
\hline
0 & 0.500 & 0.571 & 0.609 & 0.608 & 0.648 & 0.514 \\
1 & 0.599 & 0.673 & 0.608 & 0.610 & 0.620 & o \\
2 & 0.646 & 0.621 & 0.633 & 0.625 & o & o \\
3 & 0.597 & 0.631 & 0.627 & o & o & o \\
4 & 0.732 & 0.625 & o & o & o & o \\
5 & 0.431 & o & o & o & o & o \\
\hline
\hline
${\bf \gamma}$ & 0 & 1 & 2 & 3 & 4 & 5 \\
\hline
0 & 1.000 & 1.167 & 1.244 & 1.195 & 1.339 & 0.892 \\
1 & 1.200 & 1.310 & 1.213 & 1.231 & 1.230 & o \\
2 & 1.276 & 1.230 & 1.243 & 1.230 & o & o \\
3 & 1.171 & 1.242 & 1.235 & o & o & o \\
4 & 1.440 & 1.227 & o & o & o & o \\
5 & 0.845 & o & o & o & o & o \\
\hline
\end{tabular}
\end{center}

\medskip

\caption{\label{tab3}
The Pad\'e table for the values of the correlation length ($\nu$) and
the susceptibility ($\gamma$) critical exponents from the
$\varepsilon$--expansion of the pure $d=3$ Ising model.  Here and in
Tables \protect\ref{tab4}, \protect\ref{tab6}, \protect\ref{tab7} the
number of the row and of the column corresponds to the order of
numerator and denominator of a Pad\'e approximant, ``o'' means that
corresponding Pad\'e approximant cannot be constructed. }
\end{table}

\begin{table}
\begin{center}
\tabcolsep1.6mm
\begin{tabular}{||c|cccccc||}
\hline
${\bf \nu}$ & 0 & 1 & 2 & 3 & 4 & 5 \\
\hline
0 & 0.500 & 0.560 & 0.584 & 0.592 & 0.599 & 0.601 \\
1 & 0.600 & 0.699$^c$ & 0.604 & 0.495$^c$ & 0.622$^c$ & o \\
2 & 0.645 & 0.623 & 0.631$^c$ & 0.628 & o & o \\
3 & 0.597 & 0.629 & 0.629 & o & o & o \\
4 & 0.731 & 0.629 & o & o & o & o \\
5 & 0.431 & o & o & o & o & o \\
\hline
\hline
${\bf \gamma}$ & 0 & 1 & 2 & 3 & 4 & 5 \\
\hline
0 & 1.000 & 1.147 & 1.205 & 1.208 & 1.232 & 1.204$^c$ \\
1 & 1.200 & 1.359$^c$ & 1.208 & 1.205$^c$ & 1.221 & o \\
2 & 1.276 & 1.233 & 1.238$^c$ & 1.234 & o & o \\
3 & 1.171 & 1.238 & 1.236 & o & o & o \\
4 & 1.440 & 1.234 & o & o & o & o \\
5 & 0.845 & o & o & o & o & o \\
\hline
\end{tabular}
\end{center}

\medskip

\caption{\label{tab4}
The results of the Pad\'e--Borel resummation of the
correlation length ($\nu$) and the susceptibility ($\gamma$)
critical exponents from the $\varepsilon$--expansion of the pure $d=3$
Ising model. Here and in Tables \protect\ref{tab7}, \protect\ref{tab8}
the $^c$--superscript denotes that the real part of the
corresponding value is given.}
\end{table}

\begin{table}
\begin{center}
\tabcolsep1.6mm
\begin{tabular}{||c|cccccc||}
\hline
Loop & $u^*$ & $\gamma$ & $\nu$ & $\alpha$ & $\eta$ &
$\omega$ \\
\hline
2 & 0.6573 & 1.269 & 0.644 & 0.068 & 0.031 & 0.566 \\
3 & 0.4641 & 1.231 & 0.623 & 0.131 & 0.024 & 0.853 \\
4 & 0.4958 & 1.239 & 0.632 & 0.104 & 0.040 & 0.756 \\
5 & 0.4877 & 1.246 & 0.634 & 0.097 & 0.036 & 0.792 \\
\hline
\end{tabular}
\end{center}

\medskip

\caption{\label{tab5}
The results of the application of the Pad\'e--Borel resummation
($\left[ L/1\right]$) to the RG--functions of the pure $d=3$ Ising
model. Fixed point coordinate and the critical exponents of the pure
$d=3$ Ising model obtained by Pad\'e--Borel resummation in 3d scheme.
} \end{table}

\begin{table}
\begin{center}
\tabcolsep1.6mm
\begin{tabular}{||c|ccccc||}
\hline
${\bf \nu}$ & 0 & 1 & 2 & 3 & 4 \\
\hline
0 & 0.500 & 0.572 & 0.570 & 0.601 & 0.727 \\
1 & 0.601 & 0.570 & 0.572 & 0.564 & o \\
2 & 0.560 & 0.586 & 0.565 & o & o \\
3 & 0.640 & 0.541 & o & o & o \\
4 & 1.828 & o & o & o & o\\
\hline
\hline
${\bf \gamma}$ & 0 & 1 & 2 & 3 & 4 \\
\hline
0 & 1. & 1.168 & 1.140 & 1.219 & 1.783 \\
1 & 1.202 & 1.144 & 1.161 & 1.127 & o \\
2 & 1.125 & 1.172 & 1.137 & o & o \\
3 & 1.257 & 1.101 & o & o & o \\
4 & 3.824 & o & o & o & o\\
\hline
\end{tabular}
\end{center}

\medskip

\caption[]{\label{tab6}
The Pad\'e table for the values of the correlation length ($\nu$) and
the susceptibility ($\gamma$) critical exponents from the
$\sqrt{\varepsilon}$--expansion
of the $d=3$ RIM.}
\end{table}

\begin{table}
\begin{center}
\tabcolsep1.6mm
\begin{tabular}{||c|ccccc||}
\hline
${\bf \nu}$ & 0 & 1 & 2 & 3 & 4 \\
\hline
0 & 0.500 & 0.560 & 0.569 & 0.577 & 0.592 \\
1 & 0.601 & 0.573 & 0.560$^c$ & 0.565$^c$ & o \\
2 & 0.560 & 0.584 & 0.568$^c$ & o & o \\
3 & 0.639 & 0.529$^c$ & o & o & o \\
4 & 1.828 & o & o & o & o\\
\hline
\hline
${\bf \gamma}$ & 0 & 1 & 2 & 3 & 4 \\
\hline
0 & 1.000 & 1.149 & 1.148 & 1.176 & 1.252 \\
1 & 1.202 & 1.148 & 1.149 & 1.141$^c$ & o \\
2 & 1.125 & 1.168 & 1.141$^c$ & o & o \\
3 & 1.257 & 1.086$^c$ & o & o & o \\
4 & 3.825 & o & o & o & o \\
\hline
\end{tabular}
\end{center}

\medskip

\caption[]{\label{tab7}
The results of the Pad\'e--Borel resummation of the
correlation length ($\nu$) and
the susceptibility ($\gamma$) critical exponents from the
$\sqrt{\varepsilon}$--expansion
of the $d=3$ RIM.}
\end{table}

\begin{table}
\begin{center}
\tabcolsep1.6mm
\begin{tabular}{||c|cccccccc||}
\hline
Loop & $u^*$ & $v^*$ & $\gamma$ & $\nu$ & $\alpha$ & $\eta$ &
$\omega_1$ &
$\omega_2$\\
\hline
2 & 0.7886 & -0.1208 & 1.308 & 0.665 & 0.006 & 0.032 & 0.162 & 0.542
\\
3 & 0.6968 & -0.2484 & 1.293 & 0.654 & 0.039 & 0.022 & 0.430 & 0.974
\\
4 & 0.7188 & -0.1697 & 1.318 & 0.675 &-0.026 & 0.049 & 0.390$^c$ & 0.390$^c$
\\
\hline
\end{tabular}
\end{center}
\caption{\label{tab8}
Fixed point coordinates, critical exponents and stability matrix
eigenvalues
of the $d=3$ RIM obtained by Chisholm--Borel resummation in 3d scheme.}
\end{table}
\clearpage

%%%%%%%%%%%%%%%%%%%%%%%%%%%%%%
%FIGURES
%%%%%%%%%%%%%%%%%%%%%%%%%%%%%
\newpage

\newpage

\newpage

\newpage

\newpage

\begin{thebibliography}{99}
\bibitem{note1} We give here recent renormalization group estimates for the
critical exponents of the pure three dimensional Ising model obtained from
the resummation of perturbation theory series of a scalar $d=3$ $\phi^4$ theory
\cite{Guida98}.
\bibitem{note2}
Legitimacy of the replica trick as well as possible consequences of
its application will not be the subject of present study.
\bibitem{note}
However the Borel summability of the $\varepsilon$--expansion has not
been proven yet even for the model with one coupling constant.
\end{thebibliography}
\end{document}